\documentclass{aa} 

\usepackage{twoopt}
\usepackage{graphicx}
\usepackage{wrapfig}
\usepackage{float}
\usepackage{txfonts}
\usepackage{multirow}
\usepackage{tablefootnote}
\usepackage{amsmath}
\usepackage{siunitx}
\usepackage{indentfirst}
\usepackage{subfig}
\usepackage{blindtext}
\usepackage[dvipsnames]{xcolor}
\definecolor{darkgoldenrod}{rgb}{0.72, 0.53, 0.04}
\usepackage{appendix}
\usepackage{longtable}

\usepackage[colorlinks = true,
            linkcolor = RoyalBlue,
            urlcolor = darkgoldenrod,
            citecolor = RoyalBlue,
            anchorcolor = RoyalBlue, 
            draft = false]{hyperref}

\newcommand{\comment}[1]{}

\begin{document} 
   \title{Milky Way Near Twins (MWNeTs). \\
   II. Applications of a Hierarchical Framework \\
   to NGC 3521, NGC 2903, UGC 11487, and CGCG 047-026}

  \author{
I.B. Vavilova \inst{1} \thanks{Corresponding author} \and
I.O. Izviekova \inst{1,3} \email{izviekova@gmail.com}\and
A.M. Dmytrenko \inst{1} \email{admytrenko@mao.kiev.ua}\and
O.V. Kompaniiets \inst{1} \email{kompanietsolena@gmail.com}\and
O.N. Kukhar \inst{1} \email{kukhar@mao.kiev.ua}\and
O.S. Pastoven \inst{1} \email{opastoven@mao.kiev.ua}\and
S.I. Denyschenko \inst{2} \email{sofiia.denyshchenko@gmail.com}\and
D.V. Dobrycheva \inst{1} \email{daria@mao.kiev.ua}\and
P.N. Fedorov \inst{2} \email{pnfedorov@gmail.com}\and
V.P. Khramtsov \inst{2} \email{vld.khramtsov@gmail.com}\and
O. Sergijenko \inst{1} \email{olga.sergijenko.astro@gmail.com}\and
A.A. Vasylenko \inst{1} \email{vasylenko\_a@mao.kiev.ua}
}

\institute{
\inst{1} Main Astronomical Observatory of the NAS of Ukraine, 27, Akademik Zabolotnyi St., Kyiv 03143, Ukraine\\
\inst{2} Institute of Astronomy, V.N. Karazin Kharkiv National University, 35, Sumska St., Kharkiv, 61022, Ukraine \\
\inst{3} International Centre for Astronomical, Medical, and Ecological Research, National Academy of Sciences of Ukraine, 27, Akademik Zabolotnyi St., Kyiv 03123, Ukraine\\
}
\titlerunning{MWNeTs. 2. Applications}
\authorrunning{Vavilova et al.}

   \date{Received October 16, 2026; accepted , 2026}

\abstract
{Galaxies resembling the Milky Way in their present-day global properties may have experienced different evolutionary histories. The Milky Way Near Twins (MWNeT) framework introduced in Paper~I addresses this distinction through seven successive stages of present-day compatibility and evolutionary diagnostics.}
{We assess the observational application of this framework to NGC~3521, NGC~2903, UGC~11487, and CGCG~047-026, which represent different levels of observational completeness and different outcomes within the hierarchy.}
{We combine archival measurements with analyses of galaxy structure, spectral-energy-distribution similarity, atomic-gas profiles, and optical nuclear diagnostics. Stages~I--IV assess environmental, structural, nuclear, and global compatibility, while Stages~V--VII examine SEDs, rotation curves, and evolutionary fingerprints. We distinguish supported compatibility, partial agreement, measured incompatibility, and unconstrained diagnostics.}
{We present the first joint assessment of these four galaxies within the MWNeT framework and derive new structural, SED-comparison, and H\,{\sc i} profile measurements for UGC~11487. NGC~2903 provides a well-constrained benchmark, whereas NGC~3521 combines extensive evolutionary information with uncertainty in its bar identification. UGC~11487 has a securely detected bar, quantitative UV--NIR SED similarity, and distinctive transient nuclear tracers, but its nominal stellar mass lies below the adopted interval and its extended rotation curve remains unconstrained. CGCG~047-026 satisfies the initial environmental, structural, and nuclear-state criteria but remains largely unconstrained at higher stages.}
{The four applications demonstrate the framework's ability to organise complementary evidence and identify observational priorities without equating missing data with physical incompatibility. They establish a practical basis for future systematic searches and for using candidate near twins as external laboratories of Milky Way evolution.}
\keywords{Galaxy: evolution -- Galaxy: fundamental parameters -- Galaxy: structure -- galaxies: evolution -- galaxies: multiwavelength properties -- galaxies: individual: NGC~3521, UGC~11487, NGC~2903, CGCG~047-026}
\maketitle

\section{Introduction}

In Paper~I \citep{Vavilova2026}, we introduced a hierarchical framework for identifying \textit{Milky Way Near Twins} (MWNeTs): galaxies whose present-day properties resemble those of the Milky Way and whose observational signatures are consistent with broadly similar evolutionary pathways. This framework allows us to assess complementary aspects of galaxy similarity, distinguish physical incompatibility from missing observations, and identify priorities for follow-up.

The hierarchy consists of seven stages. Stage~I assesses the local environment, isolation, and cosmic-web context. Stage~II examines morphology and internal structure, including the stellar bar, disc, and bulge. Stage~III evaluates the nuclear state and supermassive black hole constraints. Stage~IV compares global spectrophotometric and dynamical properties. These first four stages provide the mandatory basis for initial candidate selection. Stage~V tests panchromatic spectral-energy-distribution (SED) similarity, Stage~VI compares resolved rotation curves, and Stage~VII examines evolutionary fingerprints preserved in chemical and stellar populations, accretion signatures, circumgalactic gas, and multiwavelength fossil tracers. Table~1 and Table~A.1 of Paper~I present the compatibility ranges and adopted Milky Way reference properties, respectively. These two tables provide the methodological and quantitative basis for the assessments presented here.

This paper applies the framework to NGC~3521, NGC~2903, UGC~11487, and CGCG~047-026. The galaxies NGC~3521 and NGC~2903 provide extensively observed systems with structural, environmental, chemical, multiwavelength, and kinematic constraints \citep{Irwin2009, McGaugh2016, Carrillo2020, Pilyugin2023, Vavilova2024, Kompaniiets2026}. UGC~11487 adds a time-domain perspective through its mid-infrared-selected tidal-disruption-event candidate \citep{Masterson2024, Izviekova2026}, while CGCG~047-026 illustrates an initial assessment based on archival photometry and spectroscopy \citep{Kruk2018}. The four galaxies constitute an illustrative sample spanning different levels of observational completeness. We combine published measurements with analyses performed in this work to assess each galaxy across the seven stages, compare the supporting evidence, and identify the observations needed to strengthen its evolutionary interpretation. 

Table~\ref{tab:mwnet_summary} summarises the stage-by-stage assessments of the four galaxies. We use four classes: \textsc{Pass}, when the available evidence satisfies the adopted criteria; \textsc{Partial}, when agreement is marginal, definition-sensitive, or limited to some diagnostics; \textsc{Fail}, when a mandatory criterion is demonstrably incompatible with the Milky Way; and \textsc{U} (unconstrained), when the observations are insufficient for a reliable assessment. Missing data therefore limit the assessment without demonstrating physical incompatibility.

Section~2 describes the common data sources and comparison methods, including diagnostic offsets relative to the Milky Way. Sections~3--6 apply the seven-stage framework to NGC~3521, NGC~2903, UGC~11487, and CGCG~047-026, respectively. Section~7 summarises the results in Table~\ref{tab:mwnet_summary}, Section~8 discusses the prospective MWNeTs' Habitable Zone, and Section~9 presents our conclusions. Appendix~A contains Table~\ref{tab:mwnet_candidates}, which compiles the measurements and constraints for the four galaxies, organised by the seven stages of the MWNeT framework. It provides the provenance of the adopted values, identifies unconstrained diagnostics, and marks quantities derived in this work as [TW].

\section{General remarks on application of the MWNeT framework to NGC~3521, UGC~11487, NGC~2903, and CGCG~047-026}
\label{sec:mwnet_applications}

\subsection{Choice of sky surveys, constraints on local environment and morphology}
\label{subsec:environment-4galaxies}

As a practical starting point for the MWNeT search, we considered existing catalogues of isolated galaxies: the Catalogue of Isolated Galaxies (CIG; $N=1051$, $z\lesssim0.03$; \citealt{Karachentseva1973}), AMIGA ($N=950$, $z\lesssim0.03$; \citealt{VerdesMontenegro2005}), 2MIG ($N=3227$, $z\lesssim0.06$; \citealt{Karachentseva2010}), 2MVIG ($N=2493$, $z\lesssim0.06$; \citealt{Karachentseva2010}), the Local Orphan Galaxies catalogue (LOG; $N=520$, $z\lesssim0.0025$; \citealt{Karachentsev2011LOG}), the SDSS-based Isolated Galaxies catalogue (SIG; $N\simeq3700$, $0.005\lesssim z\lesssim0.08$; \citealt{ArgudoFernandez2015}), and the UNAM--KIAS catalogue ($N=1520$, $0.02\lesssim z\lesssim0.08$; \citealt{HernandezToledo2010}), together with the barred-galaxy catalogue compiled by \citet{Dobrycheva2026}. These resources provide complementary samples selected using different redshift limits and isolation projected and spatial criteria.

For the initial morphological pre-selection, we retained galaxies classified as SB or SAB. This restriction is an efficient early-stage filter for a structural assessment: bar geometry, strength, deprojection, relative size, and dynamical properties are evaluated separately at Stage~II. Likewise, catalogue isolation provides the starting environmental criterion at Stage~I but does not imply the complete absence of low-mass satellites.

\begin{figure*}
\centering
\includegraphics[width=0.92\textwidth,height=0.58\textheight,keepaspectratio]{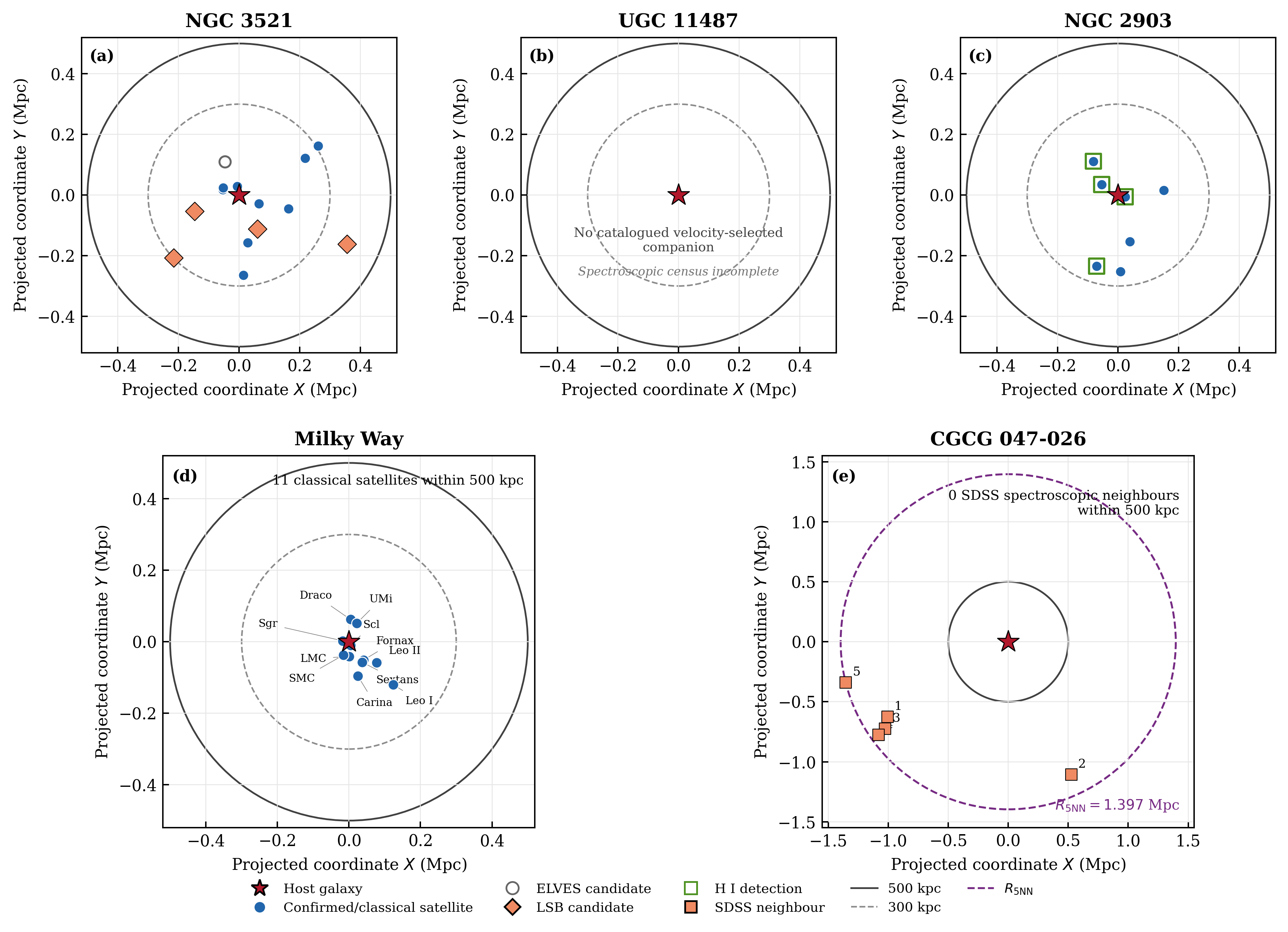}
\caption{Environments of the Milky Way and four MWNeT candidates. Solid and grey dashed circles mark projected radii of 500 and 300~kpc. Blue circles show classical MW and confirmed ELVES satellites; the open circle denotes an unconfirmed ELVES candidate. Orange diamonds mark additional low-surface-brightness candidates around NGC~3521, and green squares show H\,{\sc i}-detected satellites of NGC~2903. The purple dashed circle around CGCG~047-026 marks the fifth-nearest velocity-selected spectroscopic neighbour.}
\label{fig:stage1_environments}
\end{figure*}

We constructed the environmental maps in Fig.~\ref{fig:stage1_environments} from heterogeneous but explicitly identified data sets. For NGC~3521, we selected confirmed satellites and the previously reported candidate from the machine-readable ELVES catalogue \citep{Carlsten2022}, and we took additional low-surface-brightness candidates from Table~A.2 of \citet{Muller2025}. For NGC~2903, we selected the seven ELVES objects flagged as confirmed satellites, four of which have H\,{\sc i} detections reported by \citet{Irwin2009}.

For UGC~11487, we searched SIMBAD within $R_{\rm p}<500$~kpc and $|\Delta V|<500~{\rm km\,s^{-1}}$. At the adopted distance, these limits correspond to an angular radius of approximately $21.7\arcmin$ and a redshift interval of $0.01728<z<0.02062$. No galaxy with a catalogued redshift satisfying both limits was identified. This supports the isolated status of UGC~11487 but does not exclude faint or low-surface-brightness satellites because the available spectroscopic census is incomplete.

For CGCG~047-026, we performed an SDSS spectroscopic search using the same relative-velocity threshold, $|\Delta V|<500~{\rm km\,s^{-1}}$, and found no spectroscopically confirmed neighbour within 500~kpc. Its five nearest velocity-selected SDSS neighbours lie at projected separations of approximately $1.19$--$1.40$~Mpc, with $R_{5{\rm NN}}=1.397$~Mpc. The CGCG~047-026 panel therefore covers a wider physical scale than the other panels in order to display the fifth-nearest-neighbour environment.

\begin{figure*}
\centering
\includegraphics[width=0.88\textwidth]{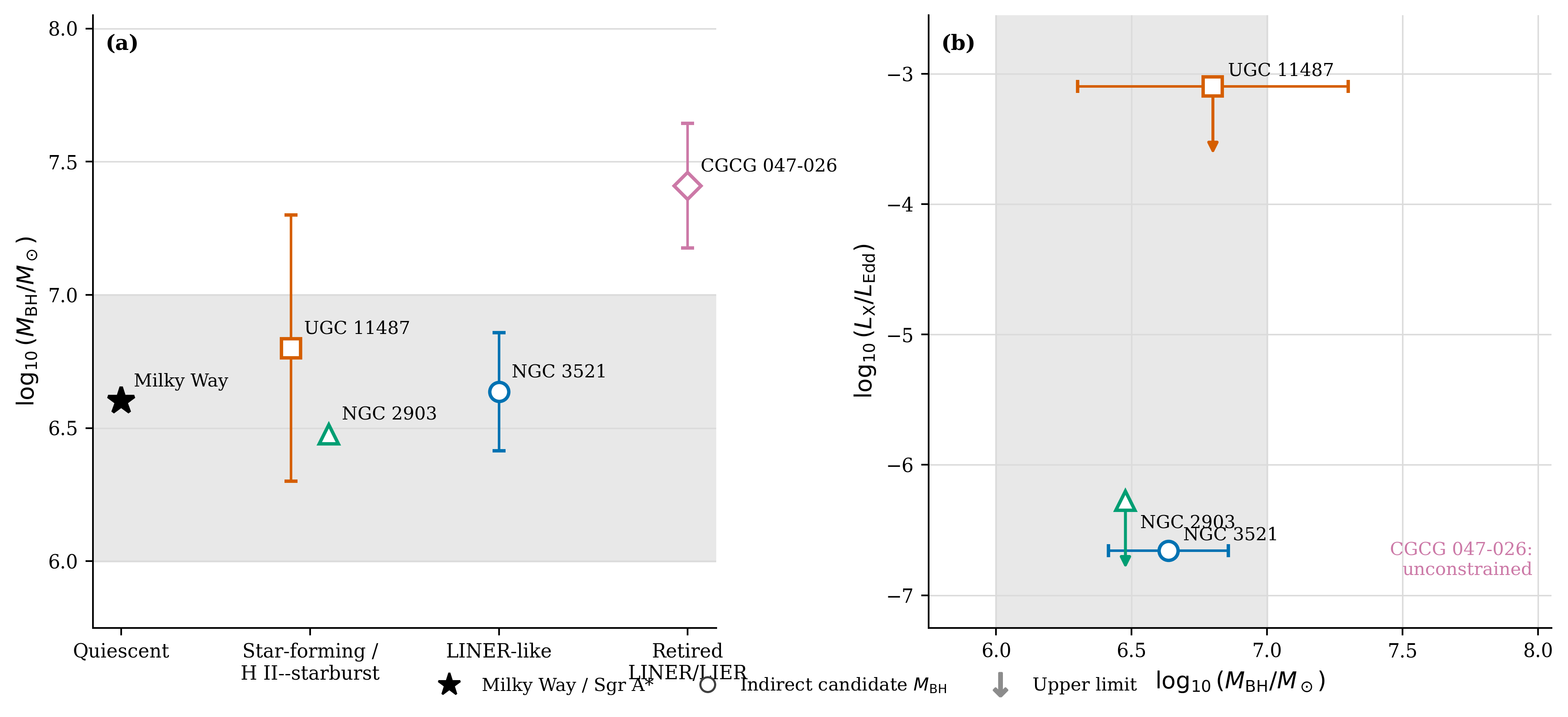}
\caption{Nuclear diagnostics of four MWNeT candidates. (a) Spectral class versus indirectly estimated black-hole mass; error bars show adopted ranges where available. The grey band marks $10^{6}\leq M_{\rm BH}/M_\odot\leq10^{7}$, and the black star denotes the Milky Way. Spectral classes are descriptive, not an evolutionary sequence. (b) X-ray-to-Eddington ratios; downward arrows indicate upper limits. CGCG~047-026 lacks a reliable nuclear luminosity constraint.}
\label{fig:app_nuclear_states}
\end{figure*}

For the external galaxies, we transformed catalogue coordinates into projected offsets relative to the corresponding host. The NGC~3521, UGC~11487, and NGC~2903 panels, together with the Milky Way panel, extend to approximately 500~kpc, whereas the CGCG~047-026 panel covers a wider field to include its fifth-nearest neighbour. We project the three-dimensional Galactocentric positions of the Milky Way satellites onto the Galactic plane, while the external-galaxy panels show projected sky-plane separations. Because the contributing surveys differ in depth, sky coverage, and spectroscopic completeness, Fig.~\ref{fig:stage1_environments} represents the currently available environmental constraints. 

\subsection{The Milky Way compatible nuclear state}
\label{subsec:nuclear-4galaxies}

At Stage~III, the Milky-Way-compatible nuclear-state set, $\mathcal{N}_{\rm MW}$, includes quiescent, star-forming, composite, LINER-like, radiatively inefficient, and weak or intermittent nuclei, provided that no energetically dominant persistent AGN is present. We treat nuclear spectral classification, black-hole mass, and accretion-state indicators as complementary diagnostics rather than requiring every candidate to exhibit the same formal optical classification as the Milky Way. 

We assess the nuclear state using three complementary diagnostics. First, optical emission-line ratios provide a spectral classification through the BPT diagrams \citep{Kewley2006}.  The complementary Veilleux--Osterbrock (VO) diagrams compare [O\,{\sc iii}]$\lambda5007$/H$\beta$ with [S\,{\sc ii}]$\lambda\lambda6717,6731$/H$\alpha$ and [O\,{\sc i}]$\lambda6300$/H$\alpha$ \citep{Veilleux1987}. Where the required lines are available, these diagnostics help distinguish H\,{\sc ii}-like, Seyfert, and LINER-like excitation, using the classification boundaries of \citet{Kewley2006}. Where the required measurements are available, the WHAN diagnostic additionally uses the H$\alpha$ equivalent width to distinguish weak AGN from retired systems whose line emission can be powered by evolved stars \citep{CidFernandes2011}. A LINER-like classification alone is therefore insufficient to establish ongoing black-hole accretion.

Second, we compare the available black-hole-mass estimates with the adopted Milky Way reference range. For these candidates, the estimates are indirect and rely on relations involving stellar velocity dispersion, bulge mass, or total stellar mass \citep{Gultekin2009, HaringRix2004, ReinesVolonteri2015}. Their interpretation accounts for intrinsic scatter, aperture effects, and uncertainties in the structural decomposition. Black-hole mass constrains accumulated growth and does not determine the current activity level.

Third, we use nuclear luminosities, upper limits, and evidence for persistent or transient activity to assess the accretion state \citep{Ho2008, Yuan2014}. We distinguish the bolometric Eddington
ratio, $\lambda_{\rm Edd}=L_{\rm bol}/L_{\rm Edd}$, from $L_{\rm X}/L_{\rm Edd}$, which requires a bolometric correction before it can be interpreted as $\lambda_{\rm Edd}$. Transient flares are assessed separately from persistent activity. The combined Stage~III assessment tests for a compatible nuclear regime; missing measurements remain unconstrained.

Figure~\ref{fig:app_nuclear_states} provides a diagnostic comparison of the nuclear states of the four illustrative MWNeT candidates. This comparison supplements the Stage~III classifications in Table~\ref{tab:mwnet_summary}. It is not intended as a conventional BPT diagram because homogeneous emission-line ratios are unavailable for all four galaxies. The comparison demonstrates that the Stage~III criterion cannot be represented by a single optical spectral class. Instead, it encompasses quiescent, star-forming, weak LINER/LIER-like, and intermittent nuclear states, provided that no dominant persistent Seyfert or quasar activity is present.

The left panel compares the adopted nuclear classifications and black-hole-mass estimates. The grey region marks the Milky-Way-compatible interval, $10^{6}\leq M_{\rm BH}/M_\odot\leq10^{7}$. The candidate galaxies span star-forming, H\,{\sc ii}/nuclear-starburst, LINER-like, and retired LINER/LIER classifications. These categories are descriptive and do not represent an evolutionary sequence. Except for Sgr~A$^\ast$, the plotted black-hole masses are indirect estimates from scaling relations and should not be interpreted as dynamical measurements. The right panel shows the available X-ray-to-Eddington ratios, $L_{\rm X}/L_{\rm Edd}$. The downward arrows for UGC~11487 and NGC~2903 denote upper limits rather than measured accretion rates. NGC~3521 has a directly reported very low X-ray-to-Eddington ratio, whereas no sufficiently reliable nuclear X-ray or bolometric luminosity is currently available for CGCG~047-026. We use these heterogeneous constraints to assess the evidence for energetically dominant persistent nuclear activity.

\subsection{Global properties: diagnostic offsets relative to the Milky Way }
\label{subsec:global-4galaxies}

\begin{figure*}[t]
\centering
\includegraphics[width=0.88\textwidth]{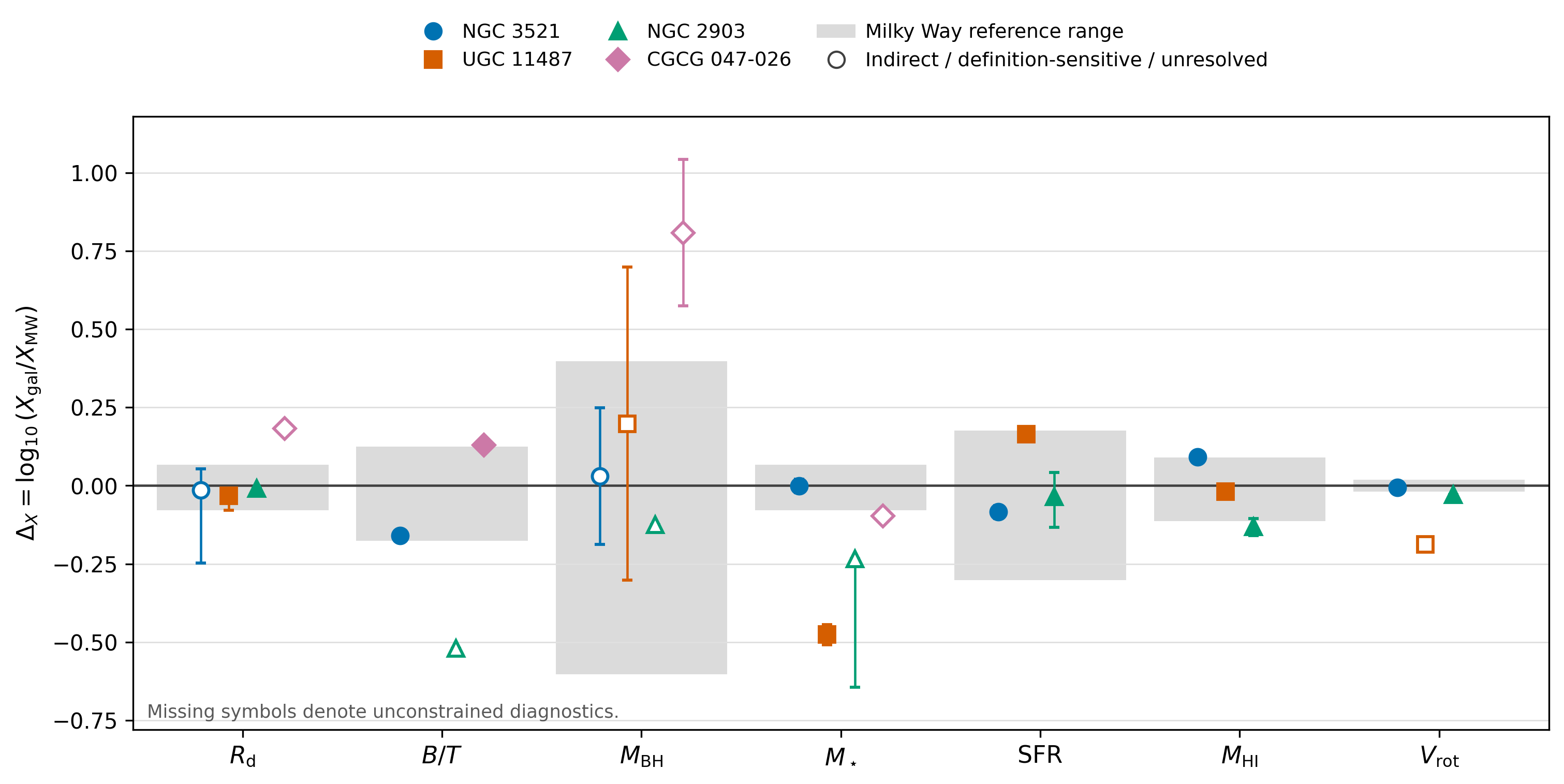}
\caption{Offsets from Milky Way fiducial values, $\Delta_X=\log_{10}(X_{\rm gal}/X_{\rm MW})$, for four candidate galaxies. Grey regions show MW reference intervals. Filled symbols denote directly comparable measurements; open symbols mark indirect, definition-sensitive, or unresolved estimates. Error bars show measurement uncertainties or literature/definition-dependent ranges. Missing symbols indicate unconstrained diagnostics.}
\label{fig:mw_normalised_offsets}
\end{figure*}

Appendix~A (Table~\ref{tab:mwnet_candidates}) presents the detailed global properties of the four illustrative candidates, while Fig.~\ref{fig:mw_normalised_offsets} compares their Milky-Way-normalised offsets. We calculated logarithmic offsets from the adopted Milky Way fiducial values as
\begin{equation}
\Delta_X=\log_{10}\left(\frac{X_{\rm gal}}{X_{\rm MW}}\right).
\end{equation}
We considered seven diagnostics for which measurements are available for at least two candidates: the stellar-disc scale length, bulge-to-total ratio, black-hole mass, total stellar mass, star-formation rate, neutral-hydrogen mass, and characteristic rotation velocity. The adopted Milky Way fiducial values are $R_{\rm d}=3.0$~kpc, $B/T=0.15$, $M_{\rm BH}=4.0\times10^{6}\,M_\odot$, $M_\star=6.0\times10^{10}\,M_\odot$, ${\rm SFR}=2.0\,M_\odot\,{\rm yr}^{-1}$, $M_{\rm HI}=6.5\times10^{9}\,M_\odot$, and $V_{\rm rot}=230\,{\rm km\,s^{-1}}$. The corresponding Milky Way reference intervals are those adopted throughout the MWNeT analysis (Paper I).

NGC~3521 broadly agrees with the Milky Way across the available global diagnostics, although its disc scale length is definition-sensitive. NGC~2903 is close to the Milky Way in disc scale length, black-hole mass, star-formation rate, H\,{\sc i} mass, and rotation velocity, but has a substantially smaller $B/T$ and a model-dependent stellar mass. UGC~11487 has a Milky-Way-compatible disc scale length, black-hole mass, star-formation rate, and H\,{\sc i} mass, while its stellar mass and unresolved linewidth-based velocity proxy lie below the adopted Milky Way values. CGCG~047-026 is compatible in $B/T$ and stellar mass but has a larger disc scale length and a higher indirect black-hole-mass estimate; several gas and kinematic diagnostics remain unconstrained. The figure therefore illustrates both each candidate's relative proximity to the Milky Way and the heterogeneous completeness of the available measurements.

\subsection{SED comparison }
\label{subsec:SED-4galaxies}
Following \citet{Kompaniiets2026}, we use broad-band SED shape as an additional diagnostic in the search for MWNeTs. Similar global SEDs may indicate broadly comparable stellar populations, attenuation, and dust-emission properties. However, SED-shape agreement alone is not sufficient to establish a physical or evolutionary analogy with the Milky Way and must therefore be considered together with independent morphological, dynamical, environmental, and multiwavelength diagnostics.

To quantify a candidate galaxy's similarity to the Milky Way, we compare its CIGALE best-fit SED with the intrinsic luminosity SED of the Milky Way \citep{Natale2022}.In this approach, both SEDs are expressed as $\lambda L_{\lambda}$ and interpolated onto a common logarithmic wavelength grid. For the Milky Way template, the tabulated $L_{\nu}$ values are converted according to $\lambda L_{\lambda} = \nu L_{\nu}$, whereas for the candidate galaxy $\lambda L_{\lambda}$ is calculated directly from the CIGALE best-fitting luminosity density $L_{\lambda}$.

\begin{figure}[t]
\centering
\includegraphics[width=\columnwidth]{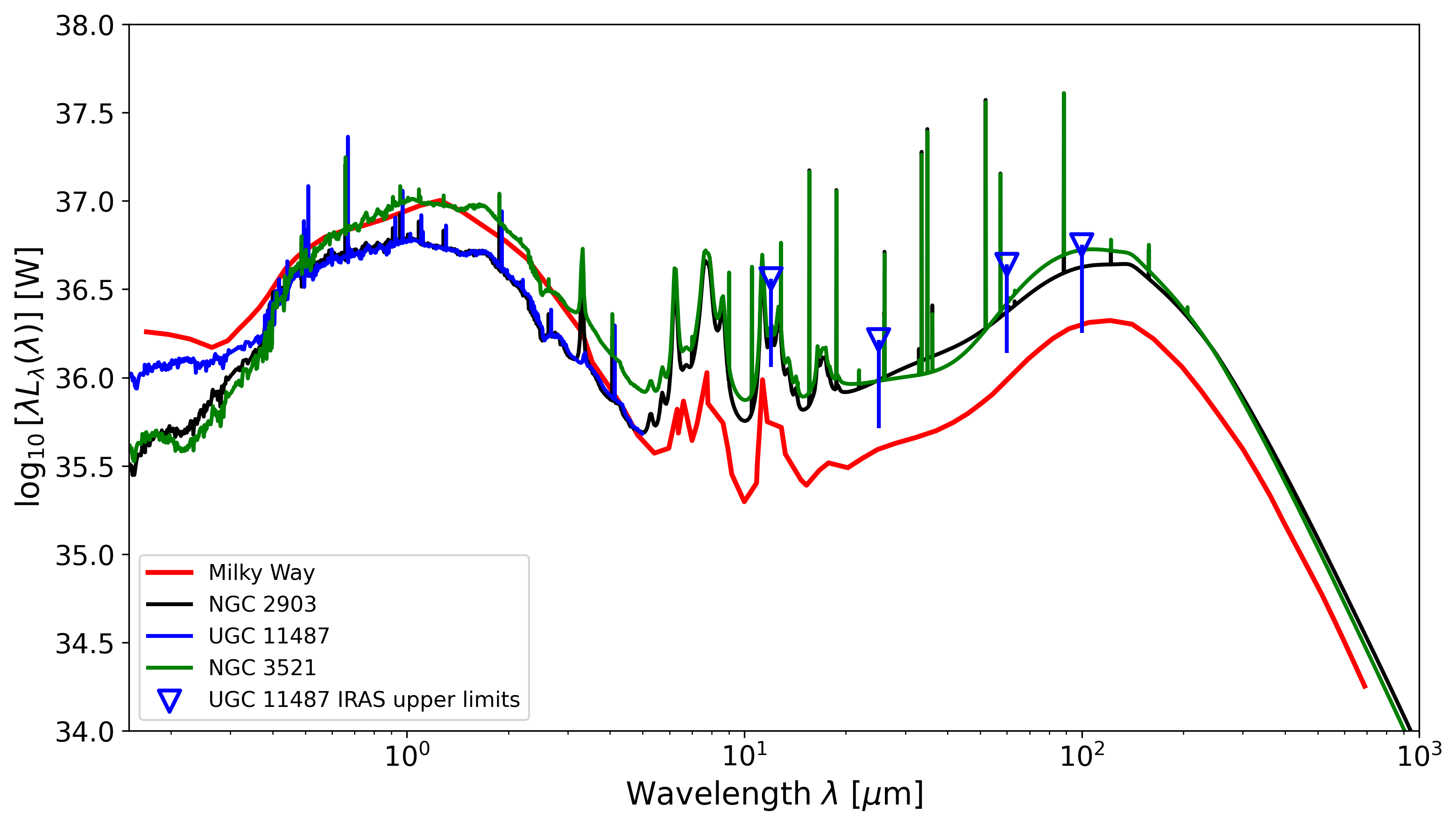}
\caption{Comparison of the intrinsic Milky Way luminosity SED with the modelled host-galaxy SEDs of NGC~2903, UGC~11487, and NGC~3521. For UGC~11487, the quantitative comparison is restricted to the common rest-frame interval $0.12$--$5\,\mu{\rm m}$; the model at longer wavelengths is shown only as a model-dependent extension, with the adopted IRAS upper limits retained as constraints.}
\label{fig:SED_MW_UGC}
\end{figure}

To compare the SED shapes independently of their absolute luminosity normalisation (Fig. \ref{fig:SED_MW_UGC}), we use the local logarithmic slope, $s(\lambda)=d\log_{10}(\lambda L_{\lambda})/d\log_{10}\lambda$. Before evaluating the derivative, we smooth the interpolated $\log_{10}(\lambda L_{\lambda})$ curves with a Savitzky--Golay filter to suppress narrow spectral features and numerical fluctuations. The slope difference between the candidate galaxy and the Milky Way is then defined as $\Delta s(\lambda)=s_{\rm gal}(\lambda)-s_{\rm MW}(\lambda)$. 

We characterise the similarity using the Pearson and Spearman correlation coefficients between the two slope sequences, along with the median slope difference, median absolute slope difference, RMS slope difference, and the fraction of wavelength-grid points satisfying a specified $|\Delta s|$ threshold. 

\citet{Kompaniiets2026} presented detailed SED measurements and modelling for NGC~3521 from the UV to decameter radio wavelengths, and \citet{Izviekova2026} do the same for UGC~11487 from the optical to the infrared. The SED analyses of NGC~2903 (Fig. \ref{fig:SED_NGC2903}) and UGC~11487 are discussed in Sections~\ref{sec:ngc2903} and \ref{sec:UGC11487}, respectively. The available data are insufficient to construct a comparable SED for CGCG~047-026.

\subsection{Rotation curves comparison}
\label{subsec:rotation_curves}

The shape and amplitude of the rotation curve provide an additional diagnostic for assessing a candidate galaxy's similarity to the Milky Way. Although the rotation curve is not a standalone criterion in the hierarchical classification, we use its overall morphology as a first check for candidates with spatially resolved kinematic measurements.

\begin{figure}
\centering
\includegraphics[width=\columnwidth]{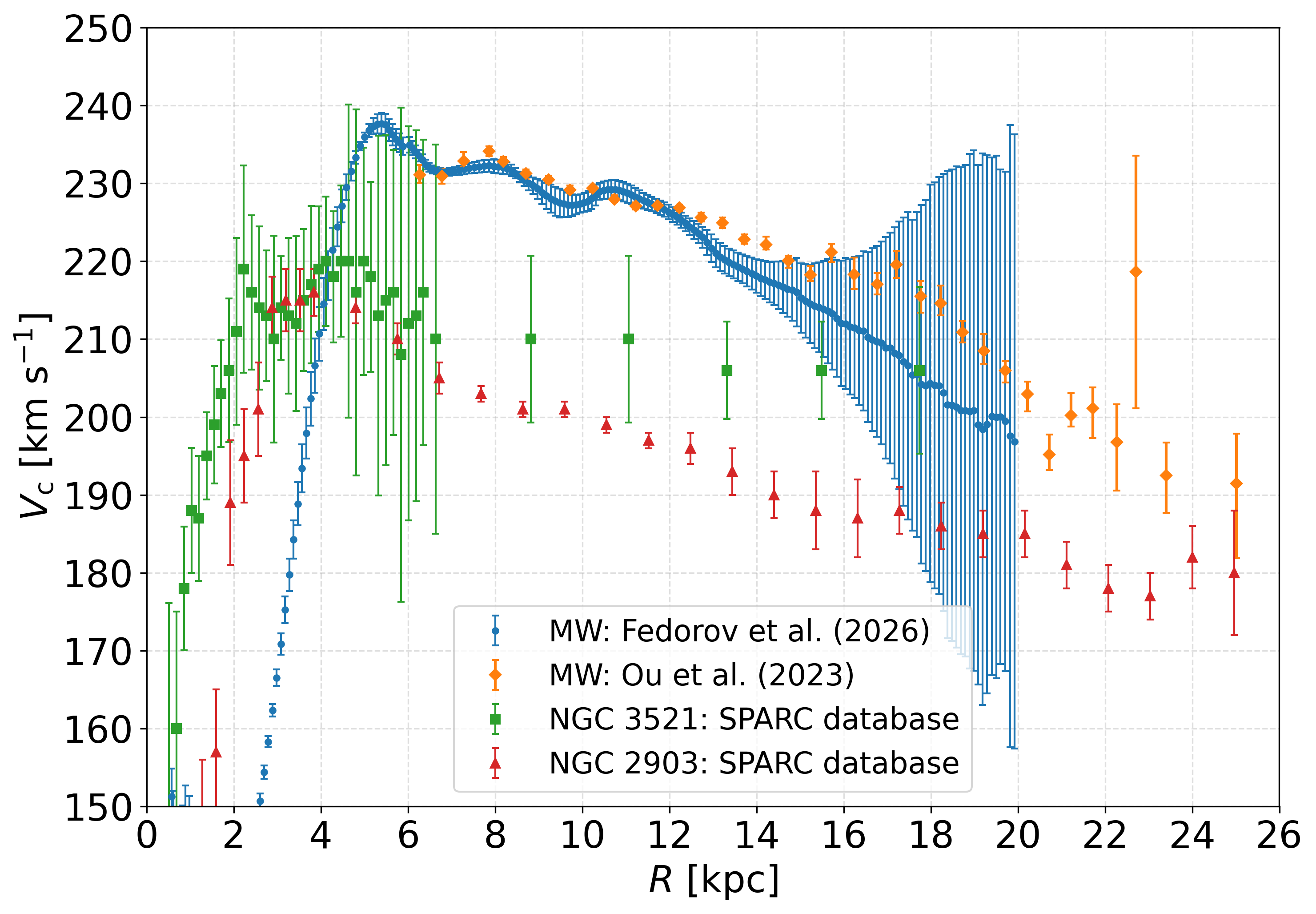}
\caption{Comparison of two Milky Way circular-velocity curves with the observed rotation curves of NGC~3521 and NGC~2903. The first Milky Way curve corresponds to the $\Theta=180^\circ$ direction and an exponential tracer-density prescription \citep{Fedorov2026}; the second Milky Way curve is adopted from \citet{Ou2024}. We use external-galaxy data from the SPARC compilation \citep{Lelli2016}, with NGC~3521 measurements from the THINGS analysis of \citet{deBlok2008}. The external-galaxy curves represent deprojected observed rotation velocities and may include residual non-circular motions.}
\label{fig:rotation_curves_comparison}
\end{figure}

Figure~\ref{fig:rotation_curves_comparison} compares two independent Milky Way circular-velocity curves with the resolved rotation curves of NGC~3521 and NGC~2903. The first Milky Way curve was derived from RGB stars in Gaia~DR3 using the radial Jeans equation and an exponential model for the tracer-density distribution \citep{Fedorov2026}. We adopt the curve at azimuth $\Theta=180^\circ$, which provides a representative profile. We take the second Milky Way curve from \citet{Ou2024}, who used 120,309 stars with APOGEE~DR17 spectroscopy combined with Gaia~DR3, 2MASS, and WISE photometry. Their analysis extends to approximately 30~kpc and indicates a more pronounced decline at large Galactocentric radii than in the inner Galaxy.

The SPARC data provide observed rotation velocities and their uncertainties as a function of galactocentric radius \citep{Lelli2016}. These measurements are commonly used to characterise disc-galaxy rotation curves. Nevertheless, residual non-circular motions, inclination uncertainties, warps, and kinematically distinct gas components can affect the inferred curves. This issue is particularly relevant to NGC~3521 because its outer H,{\sc i} distribution contains a diffuse, slowly rotating anomalous component \citep{Elson2014}. The present comparison should therefore be interpreted primarily in terms of global curve morphology and characteristic velocity scale, rather than as a direct comparison of identical dynamical quantities.

The four curves occupy broadly comparable velocity ranges, with maximum velocities between approximately 216 and 238~km~s$^{-1}$. The Milky Way curves of \citet{Fedorov2026} and \citet{Ou2024} reach maximum values of approximately 238 and 234~km~s$^{-1}$, respectively. The corresponding observed maximum rotation velocities of NGC~3521 and NGC~2903 are approximately 220 and 216~km~s$^{-1}$. Thus, both candidates are consistent with the Milky Way's characteristic velocity scale.

To quantify the radial decline, we fitted a linear relation, $V(R)=a+bR$, over the common radial interval $6\leq R\leq20$~kpc. For each galaxy, only data points within this interval were included in the fit. We determined the slope $b$ using ordinary unweighted least-squares regression, which describes the mean radial trend over the selected interval. The Milky Way curves have slopes of approximately $-1.9$ and $-1.81$~km~s$^{-1}$~kpc$^{-1}$ for \citet{Fedorov2026} and \citet{Ou2024}, respectively.  For the external galaxies, we obtain $-1.70\pm0.08$~km~s$^{-1}$~kpc$^{-1}$ for NGC~2903 and $-0.68\pm0.15$~km~s$^{-1}$~kpc$^{-1}$ for NGC~3521. Thus, NGC~2903 shows a radial decline similar to that of the Milky Way, whereas NGC~3521 has a substantially flatter outer rotation curve.

The comparison remains sensitive to the adopted tracer population, density prescription, radial coverage, and dynamical assumptions. In addition, the outer measurements of NGC~3521 and the Milky Way curve derived here have comparatively larger uncertainties. We therefore treat the rotation-curve morphology as a complementary diagnostic within the MWNeT framework. NGC~3521 is consistent with the Milky Way in characteristic rotation amplitude and, relative to the Fedorov et al. curve, in the approximate radius of maximum velocity, but its outer rotation curve declines more slowly. Consequently, its rotation-curve diagnostic is classified as \textsc{Partial}, while its characteristic velocity remains consistent with the Milky Way range.

\section{NGC~3521: a strong MWNeT pathfinder candidate}
\label{subsec:NGC3521}

NGC~3521 is a nearby isolated flocculent spiral and one of the most extensively observed Milky Way analogues. Its similarity to the Milky Way was first recognised from its dynamics and structure. \citet{McGaugh2016} identified NGC~3521 as a near twin based on its luminosity, disc scale length, and rotation curve, while \citet{Pilyugin2023} later classified it as both a structural and an evolutionary analogue using its stellar mass, optical size, rotation velocity, and radial oxygen-abundance distribution. The authors have also examined its multiwavelength properties in previous studies \citep{Pastoven2024, Kompaniiets2026}. Table~\ref {tab:mwnet_candidates} compiles the principal measurements and references making it particularly suitable for testing the MWNeT hierarchy.

\textit{isolation and cosmic-web context.} NGC~3521 is identified as KIG~461 and satisfies the classical Karachentseva isolation criterion \citep{Karachentseva1973}. Its local density and location within the diffuse Leo Spur broadly match the environmental regime adopted for the Milky Way \citep{Kourkchi2017, Kompaniiets2025, Vavilova2024, Kompaniiets2026}. The reported dwarf satellites and additional low-surface-brightness candidates do not contradict this criterion, which excludes dynamically significant comparable-mass neighbours \citep{Carlsten2022, Muller2025}. We therefore assign a \textsc{pass} classification at Stage~I (see Fig. \ref{fig:stage1_environments}).

\textit{structural and morphological constraints.} The detailed morphology of NGC~3521 is catalogue-dependent: RC3 classifies it as SAB, whereas the CVRHS/S$^4$G classification is formally unbarred \citep{deVaucouleurs1991RC3, Buta2015Cat}. A weak, nearly end-on bar has also been proposed from the asymmetric central stellar kinematics, but it has not been established unambiguously \citep{Zeilinger2001, Coccato2018}.

At the high inclination of NGC~3521, a Milky Way-like bar could be foreshortened, obscured by the bulge, disc, and dust, or blended with the inner ring and spiral structure. Thus, the lack of an unambiguous photometric detection does not demonstrate the physical absence of a bar. Because the initial morphological pre-selection explicitly admitted both SB and SAB system (Sect.~\ref{sec:summary_assessment}), NGC~3521 receives a \textsc{pass} classification at Stage~II, with the important caveat that the physical identification and deprojection of its putative bar remain uncertain.

\textit{Milky-Way-compatible nuclear state.} The nucleus is classified as H\,{\sc ii}/LINER or LINER-like, depending on the adopted diagnostic and aperture \citep{Kompaniiets2025, Kompaniiets2026}. These line ratios alone do not establish a persistent radiatively efficient AGN. Independent scaling relations place the central black hole in the same low-mass regime as Sgr~A$^*$, while its exceptionally low X-ray-to-Eddington luminosity ratio argues against a presently dominant AGN \citep{Zhang2009, Davis2014}. NGC~3521 therefore receives a \textsc{pass} classification at Stage~III.

\textit{Global physical constraints.} The stellar mass, luminosity, star-formation rate, dust luminosity, optical size, rotation velocity, and atomic- and molecular-gas contents are broadly compatible with the adopted Milky Way ranges (Table~\ref{tab:mwnet_candidates}). For consistency, we calculated the baryonic gas fraction as $f_{\rm gas}=1.36(M_{\rm HI}+M_{\rm H_2})/[M_\star+1.36(M_{\rm HI}+M_{\rm H_2})]$, where the factor 1.36 accounts for helium. The tabulated quantities give $f_{\rm gas}=0.195$.

The inferred halo mass remains strongly dependent on the assumed density profile, as illustrated by the different NFW and Burkert results in Table~\ref{tab:mwnet_candidates}. We therefore use it as a supporting rather than decisive diagnostic. Overall, NGC~3521 receives a \textsc{pass} classification at Stage~IV.

\textit{Panchromatic SED similarity.} The integrated SED broadly resembles that of the Milky Way from UV to radio/decametre wavelengths with moderate residual differences in the dust-emission and centimetre-radio components \citet{Kompaniiets2026}. NGC~3521 receives a \textsc{pass} classification at Stage~V.

\textit{Rotation-curve similarity.} NGC~3521 agrees with the Milky Way in characteristic rotation amplitude and approximately in the radius of maximum velocity, but its outer curve declines substantially more slowly. Section~\ref{subsec:rotation_curves} presents the quantitative comparison. Accordingly, we assign a \textsc{partial} classification at Stage~VI.

\textit{Evolutionary fingerprints and fossil tracers.} The broken abundance gradient and effective oxygen yield are consistent with inside-out disc evolution without requiring strong present-day net gas exchange \citep{Pilyugin2025}. Resolved stellar populations indicate old, intermediate-age, and young components, although the intermediate-age population does not uniquely distinguish secular evolution from minor accretion \citep{Coccato2018}. The diffuse stellar envelope and possible shell-like substructures provide additional, though not unique, evidence for past minor accretion \citep{MartnezDelgado2025}.

The evolutionary diagnostics of NGC~3521 also provide an observational basis for future modelling of its Galactic Habitable Zone (GHZ). In particular, the radial chemical-abundance distribution, gas content, star formation properties, stellar populations, and disc dynamics offer complementary constraints on the spatial and temporal evolution of astrophysical conditions potentially relevant to planetary habitability. Their combination may enable the development of chemo-dynamical GHZ models, including the effects of chemical enrichment, energetic events, and stellar orbital evolution. Such modelling lies beyond the scope of the present study and represents a prospective application of the MWNeT evolutionary framework (see Section \ref{sec:mwnets_habitable_zone}).

\textit{Integrated MWNeT assessment.} NGC~3521 passes the isolation, catalogue-based morphological, nuclear-state, global-property, SED, and evolutionary-fingerprint stages. Its rotation amplitude is Milky Way-like, although the flatter outer decline yields a \textsc{partial} rotation-curve assessment. The agreement across these independent diagnostics identifies NGC~3521 as a strong MWNeT pathfinder candidate. Its principal qualifications are the uncertain physical identification of the bar and the flatter outer rotation curve.

\section{NGC~2903: a strong MWNeT pathfinder candidate}
\label{sec:ngc2903}

\textit{Stages I--IV, VII.} NGC~2903 provides one of the most extensively constrained applications of the MWNeT hierarchy considered in this work. It combines a catalogue-based isolated status with a well-resolved barred morphology, a Milky-Way-compatible nuclear state, extensive multiwavelength coverage, resolved gas and stellar-population measurements, and a spatially resolved rotation curve. 
Table~\ref{tab:mwnet_candidates} summarises the corresponding SED diagnostics for the four illustrative candidates. Here we discuss only their combined interpretation and our additional results.

At Stage~I, NGC~2903 satisfies the adopted catalogue-based isolation criterion. The presence of low-mass satellites does not invalidate this classification, because the MWNeT isolation criterion excludes recent dynamically important interactions and comparable-mass companions rather than requiring a completely empty satellite environment. The currently available environmental evidence therefore supports a relatively quiescent evolutionary setting broadly compatible with that of the Milky Way (see Fig. \ref{fig:stage1_environments}).

NGC~2903 also passes the Stage~II structural assessment. Its strong stellar bar is detected directly in optical and near-infrared imaging, and its disc scale length and deprojected bar size fall within the Milky-Way-compatible ranges adopted in the framework. The measured corotation-to-bar ratio is consistent with a dynamically fast bar. The comparatively small bulge-to-total ratio indicates a particularly disc-dominated system and qualifies the detailed structural comparison, but does not demonstrate incompatibility with the broader barred-disc criterion.

At Stage~III, the nuclear spectrum is dominated by H\,{\sc ii}-region or nuclear-starburst excitation. Although a very weak accreting component cannot be excluded, there is no evidence for a dominant persistent AGN. The estimated central black-hole mass is also close to the mass scale of Sgr~A*. NGC~2903 is therefore consistent with the broad Milky-Way-compatible nuclear-state criterion adopted in Paper I. 

The global stellar, star-formation, gas, dust, size, and dynamical properties listed in Table~\ref{tab:mwnet_candidates} broadly agree with the Milky Way at Stage~IV when the literature stellar-mass estimate of $3.50\times10^{10}\,M_\odot$ is adopted. Our homogeneous SED fit gives a lower value of $(1.36\pm0.21)\times10^{10}\,M_\odot$, below the adopted interval, so the stellar-mass compatibility remains model-dependent. The combined assessment is supported by the size, star-formation, gas-content, and characteristic-velocity diagnostics, but the mass discrepancy qualifies the Stage~IV classification. Available H\,{\sc i} and molecular-gas measurements show a substantial cold-gas reservoir capable of sustaining its current star formation. Fig.~\ref{fig:mw_normalised_offsets} presents its logarithmic offsets from the adopted Milky Way values.

At Stage~VII, several independent diagnostics constrain its evolutionary history. The resolved abundance gradient is consistent with sustained radial chemical evolution of the disc. The old, metal-rich central population and the intermediate-age populations associated with the bar indicate a combination of early central assembly and subsequent bar-driven secular evolution. The available stellar-population analysis does not support a recent major merger; instead, it favours growth through minor mergers, gas accretion, and internal secular processes. Diffuse soft X-ray-emitting gas extending beyond the central star-forming region provides additional evidence for starburst-driven disc--halo feedback. Its morphology does not, however, establish an unambiguous large-scale galactic wind, and neither the extended circumgalactic medium nor the globular-cluster system is currently constrained homogeneously enough for a detailed MW comparison.

\begin{figure}[t]
\centering
\includegraphics[width=\columnwidth]{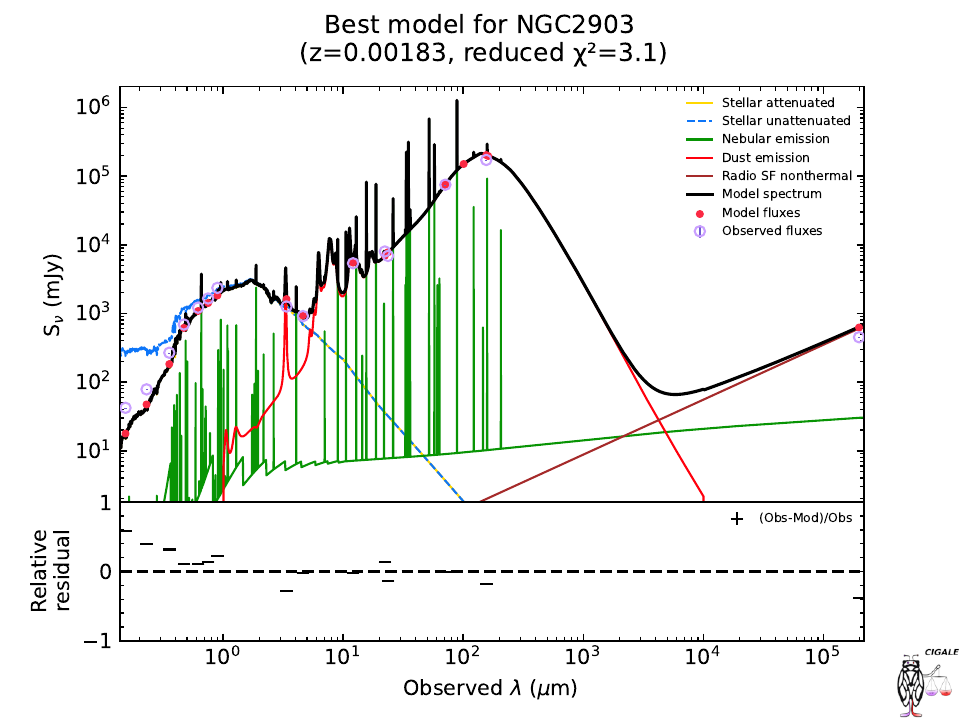}
\caption{Best-fitting CIGALE model of the integrated SED of NGC~2903. The derived global parameters are listed in Table~\ref{tab:mwnet_candidates}.}
\label{fig:SED_NGC2903}
\end{figure}

\textit{Panchromatic SED and rotation curve comparisons.} We performed multiwavelength aperture photometry and SED modelling of NGC~2903 following the methodology applied to NGC~3521 by \citet{Kompaniiets2026}, with minor adjustments to the CIGALE parameter grid. Because the available radio coverage extends only to the VLA L~band, we employed the standard CIGALE \texttt{radio} module. The resulting stellar, star-formation, dust, and luminosity parameters are reported in Table~\ref{tab:mwnet_candidates} and are not repeated here. Fig.~\ref{fig:SED_NGC2903} shows the best-fit model compared with the intrinsic Milky Way luminosity SED, using the same derivative-based and normalised-shape diagnostics as for NGC~3521. Over the common UV-to-far-infrared wavelength interval, the logarithmic-slope sequences have Pearson and Spearman correlation coefficients of $r=0.870$ and $r_{\rm S}=0.850$, respectively. The median slope difference is $\widetilde{\Delta s}=0.024$, while the median absolute slope difference and the RMS difference are 0.345 and 1.237, respectively. Approximately 57\% of the common wavelength-grid points satisfy $|\Delta s|<0.5$.

The normalised SEDs provide a complementary measure that is less sensitive to the absolute luminosity scale. Their median absolute difference is 0.318~dex, and their RMS difference is 0.390~dex. Approximately 89\% of the common wavelength-grid points differ by less than 0.5~dex, while 31\% agree within 0.25~dex. The small median slope offset indicates no big systematic difference between the local SED gradients of NGC~2903 and the Milky Way. The larger absolute and RMS differences nevertheless show that the correspondence is not uniform and that appreciable deviations remain in individual spectral regions.

The integrated luminosity distribution across both the stellar-dominated and dust-emission regimes is therefore broadly similar to that of the Milky Way, without requiring point-by-point identity between the two SEDs. This panchromatic agreement provides an independent Stage~V consistency diagnostic supporting the MWNeT classification. Table~\ref{tab:mwnet_candidates}compares these results with other illustrative candidates.

The resolved rotation curve provides an additional independent consistency test. As shown in Section~\ref{subsec:rotation_curves}, NGC~2903 has a characteristic rotation amplitude comparable to that of the Milky Way. The linear fit performed in the present work over the common radial interval yields an outer-curve slope that declines, similar to those obtained for the two adopted Milky Way curves. This agreement concerns the global velocity scale and mean radial trend; it should not be interpreted as point-by-point identity between the observed external-galaxy rotation velocity and the inferred Milky Way circular velocity.

\textit{Integrated MWNeT assessment}. NGC~2903 passes the isolation, structural, nuclear-state, global-property, panchromatic, and resolved-kinematic assessments of the MWNeT hierarchy and exhibits several compatible evolutionary fingerprints. Its classification is supported by mutually independent evidence: a securely detected stellar bar, a Milky-Way-compatible central black hole mass and nuclear state, substantial cold-gas reservoirs, ongoing moderate star formation, resolved chemical and stellar-population structure, broad UV-to-far-infrared SED similarity, and a declining rotation curve with a Milky Way-like characteristic velocity scale.

The principal qualifications are its unusually small bulge-to-total ratio, the systematic spread among stellar-mass estimates, and the incomplete characterisation of its globular-cluster system and extended circumgalactic medium. These limitations represent measurement or coverage uncertainties rather than demonstrated failures of mandatory MWNeT criteria. We therefore assign NGC~2903 a \textsc{Pass} in the integrated assessment and classify it as a strong, well-constrained MWNeT pathfinder candidate.

\section{UGC~11487: a time-domain-selected MWNeT candidate}
\label{sec:ugc11487}

UGC~11487, the host of the infrared-selected TDE-like transient WTP14adeqka, is examined here for the first time within the MWNeT framework. We assess its environmental, structural, nuclear, and global properties at Stages~I--IV, followed by SED similarity, available kinematic constraints, and evolutionary indicators at Stages~V--VII. The transient provides information on recent nuclear activity and circumnuclear reprocessing, distinct from the galaxy-wide evolutionary diagnostics. Quantities derived or reassessed in this work are marked [TW] in Table~\ref{tab:mwnet_candidates}.

\textit{Isolation and environmental context.} UGC~11487 belongs to the 2MIG catalogue of isolated galaxies \citep{Karachentseva2010}. We performed a supplementary SIMBAD search for neighbours within a projected radius of 500\,kpc and a line-of-sight velocity difference of $|\Delta V|<500\,{\rm km\,s^{-1}}$. At the adopted redshift $z=0.01895$ and luminosity distance $D_{\rm L}=82.34\,{\rm Mpc}$, these limits correspond to an angular radius of approximately $21.7\,{\rm arcmin}$ and a redshift interval of $0.01728<z<0.02062$. No object with a catalogued redshift satisfying both conditions was identified (Fig.~\ref{fig:stage1_environments}). Objects without measured redshifts could not be assessed, so this search does not constitute a complete satellite census. Together with the catalogue-based isolation classification, the result supports a \textsc{Pass} assessment for the adopted low-density-environment criterion at Stage~I.

The location of UGC~11487 relative to the cosmic web remains unconstrained because it lies outside the reliable footprint of the SDSS/Bisous and SDSS/DisPerSE filament reconstructions considered here. Filament-spine and node distances are therefore unavailable. This limitation concerns the supplementary characterisation of its large-scale environment and does not alter the assessment of the local-neighbour criterion.

\begin{figure*}[t]
\centering
\includegraphics[width=0.95\textwidth]{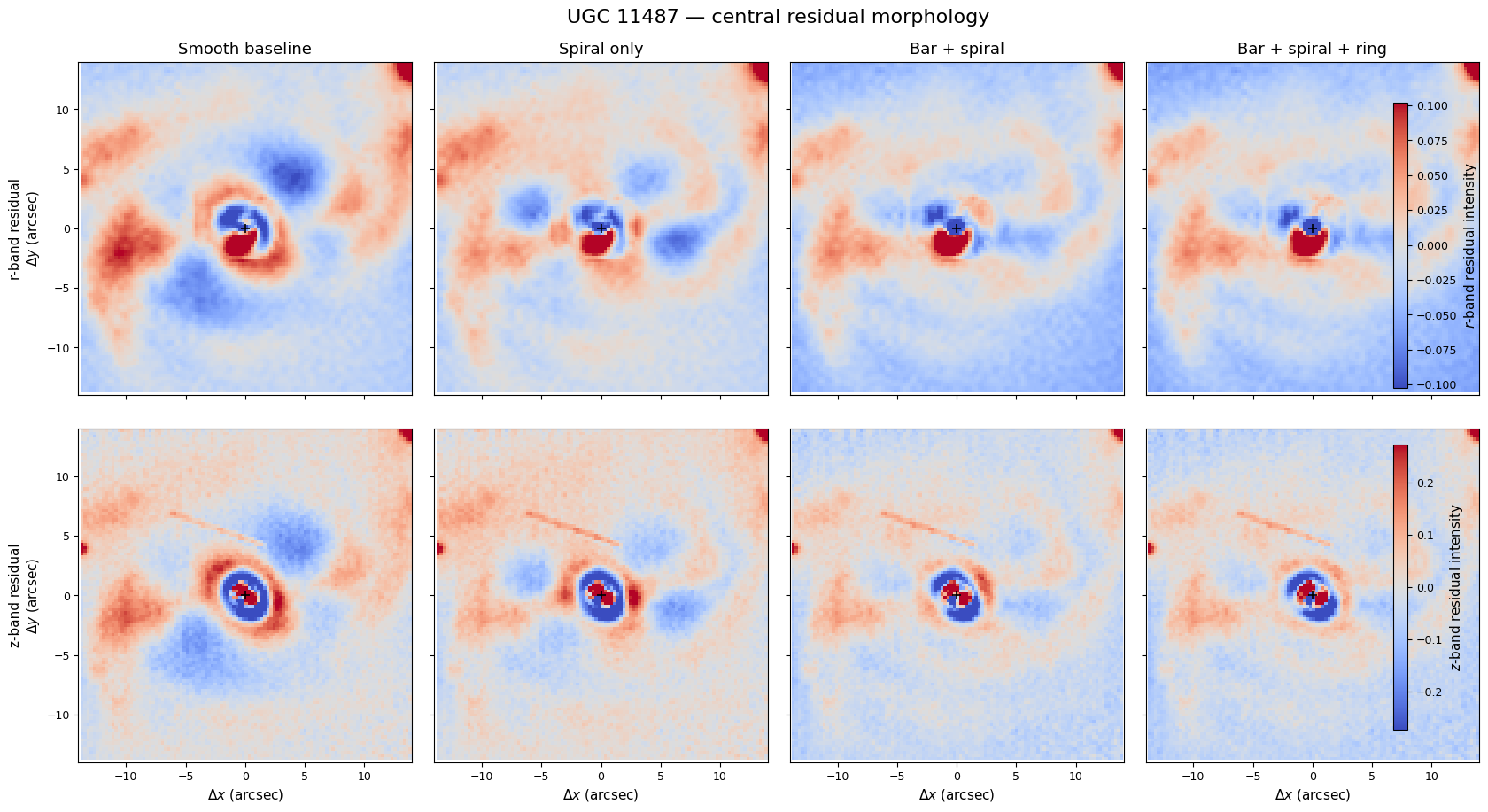}
\caption{Central residuals (data minus model) of UGC~11487 in the DESI Legacy Surveys $r$ and $z$ bands. Columns show the smooth baseline, winding $m=2$, bar-plus-spiral, and bar-plus-spiral-plus-ring models. Symmetric colour scales are set separately for each band; the ring component is not adopted.}
\label{fig:ugc11487_residuals}
\end{figure*}

\textit{Structural and morphological constraints.} UGC~11487 is classified as an SBb galaxy. We modelled DESI Legacy Imaging Surveys $r$- and $z$-band images \citep{Dey2019Legacy} using an outer exponential disc and central components representing smooth stellar emission, a winding $m=2$ structure, and a Ferrers-type bar \citep[e.g.][]{Salo2015}. The disc fit gives $h=2.78^{+0.15}_{-0.28}\,{\rm kpc}$ and an apparent axis ratio of $q_{\rm disc}=0.526^{+0.028}_{-0.023}$ [TW]. Adopting an intrinsic disc thickness of $0.13\leq q_0\leq0.20$ gives a working inclination of $i\simeq59.6\,{\rm deg}$ \citep{Tully1977,Hall2012}. The combined bar-plus-spiral model reduces the coherent central residuals in both bands (Fig.~\ref{fig:ugc11487_residuals}).

We tested the bar detection using two alternative smooth central-background prescriptions and 5000 empirical null realisations per prescription, repeating the full bar search in each realisation. The observed improvements, $\Delta{\rm BIC}_{\rm bar}=136.4$ and 316.2, exceed the corresponding 99\% null thresholds of 46.0 and 116.2. No null realisation reached the observed improvements, giving $p_{\rm bar}\simeq2.0\times10^{-4}$ for each prescription \citep{PhipsonSmyth2010}. These results support the presence of a stellar bar within the adopted model family, conditional on the fixed disc geometry, centre, sky level, and PSF.

For the adopted disc geometry, the nominal deprojected bar semi-major axis is $a_{\rm bar}\simeq4.8\,{\rm kpc}$, with an intrinsic axis ratio of $q_{\rm bar}\simeq0.43$ and $a_{\rm bar}/h\simeq1.7$ [TW]. These parameters remain provisional because uncertainties in the disc geometry and fixed fitting inputs have not been fully propagated. An additional inner ring-like component reached the lower boundary of the allowed width under both central-background prescriptions; we therefore do not adopt a statistically secure separate inner ring or pseudoring. The bulge-to-total ratio, bulge classification, corotation radius, and bar pattern speed remain unconstrained.

The disc morphology, compatible scale length, and statistically supported bar detection support Stage~II=\textsc{Pass} under the adopted disc-and-bar criteria.

\textit{Milky-Way-compatible nuclear state.} We analysed the reduced, flux-calibrated spectrum of UGC~11487 obtained with the Binospec spectrograph on the 6.5-m MMT telescope, provided by Walter Golay and originally presented by \citet{Golay2026}. The observations comprise two 600-s exposures obtained on 2024 March 17 with a $1\,{\rm arcsec}$ slit. We used the pipeline-extracted one-dimensional spectrum and modelled the stellar continuum with pPXF under alternative E-MILES and FSPS prescriptions to assess the nuclear excitation at Stage~III.

Emission lines were fitted in the continuum-subtracted spectra, with fixed theoretical flux ratios for the [O\,{\sc iii}], [O\,{\sc i}], and [N\,{\sc ii}] doublets. The H$\alpha$+[N\,{\sc ii}]+[S\,{\sc ii}] complex was modelled with two narrow kinematic components, and their summed fluxes were used for the excitation diagnostics. H$\beta$ was fitted with a single Gaussian; direct integration of its continuum-subtracted profile under both stellar-continuum prescriptions gave consistent total fluxes, supporting its use in the diagnostic line ratios.

Emission-line flux uncertainties were estimated using moving-block residual-bootstrap realisations with block lengths of 7, 9, and 11 spectral pixels, combining the E-MILES and FSPS continuum-subtraction results with equal weight. The quoted intervals span the 16th--84th percentiles. The measured fluxes are not corrected for Galactic foreground or internal attenuation.

\begin{table}[t]
\centering
\caption{Observed emission-line fluxes of UGC~11487 in units of
$10^{-17}\,\mathrm{erg\,cm^{-2}\,s^{-1}}$, without attenuation corrections.}
\label{tab:ugc11487_emission_lines}
\begin{tabular}{lc}
\hline
Line & Flux \\
\hline
H$\beta$ & $1730.8^{+52.4}_{-46.3}$ \\
{}[O\,\textsc{iii}] $\lambda5007$ & $317.4^{+16.4}_{-19.0}$ \\
{}[O\,\textsc{i}] $\lambda6300$ & $336.7^{+52.2}_{-50.6}$ \\
H$\alpha$ & $12876.1^{+139.9}_{-143.7}$ \\
{}[N\,\textsc{ii}] $\lambda6583$ & $7919.4^{+118.7}_{-111.0}$ \\
{}[S\,\textsc{ii}] $\lambda6716$ & $1977.3^{+84.1}_{-83.4}$ \\
{}[S\,\textsc{ii}] $\lambda6731$ & $2025.9^{+84.4}_{-89.7}$ \\
\hline
\end{tabular}
\par\smallskip
\begin{minipage}{\columnwidth}
\footnotesize
\raggedright
Errors give the 16th--84th percentile intervals.
H$\alpha$, [N\,\textsc{ii}], and [S\,\textsc{ii}] fluxes
sum both narrow components.
\end{minipage}
\end{table}

The emission-line ratios derived from Table~\ref{tab:ugc11487_emission_lines} place the central spectrum in the composite region of the [N\,{\sc ii}]-BPT diagram and in the star-forming regions of the [S\,{\sc ii}] and [O\,{\sc i}] diagrams, consistent with \citet{Golay2026} (Fig.~\ref{fig:app_nuclear_states}). Taken together, these diagnostics do not require a dominant persistent AGN, although a weak contribution cannot be excluded. The composite classification alone does not establish an incompatibility with the weak nuclear regime adopted for Stage~III.

Tests with an additional broad H$\alpha$ component do not yield a secure detection: the residual-bootstrap significance depends on the adopted stellar-continuum prescription. We therefore do not infer a broad-line virial black-hole mass. Together with the excitation diagnostics, this result provides no compelling evidence for a persistent broad-line AGN.

The indirect black-hole mass estimate of $\log(M_{\rm BH}/M_\odot)=6.8\pm0.5$ reported by \citet{Masterson2024} was obtained from the total host stellar mass using the relation of \citet{ReinesVolonteri2015}. We derived the host mass used in that estimate with \textsc{Prospector}, and it differs from the CIGALE estimate adopted for the global diagnostics in this work. The nominal value, $M_{\rm BH}\simeq6.3\times10^6\,M_\odot$, is consistent with the adopted Milky Way-like black-hole mass range, although uncertainties and calibration dependence prevent a precise comparison with Sgr~A$^*$.

For this nominal mass, $L_{\rm Edd}=1.26\times10^{38}(M_{\rm BH}/M_\odot)\simeq7.9\times10^{44}\,{\rm erg\,s^{-1}}$. Late-time Swift/XRT observations yielded no detection, with a reported upper limit of $L_{\rm X}(0.3$--$10\,{\rm keV})\lesssim6.3\times10^{41}\,{\rm erg\,s^{-1}}$ \citep{Masterson2024, Golay2026}. This gives $L_{\rm X}/L_{\rm Edd}\lesssim8\times10^{-4}$ [TW]. The limit applies to the observed epoch and energy band; a bolometric Eddington ratio cannot be inferred without an adopted bolometric correction.

Taken together, the available diagnostics support Stage~III=\textsc{Pass} under the adopted weak-nuclear-activity criterion. The TDE-like flare is treated as an episodic event and does not establish persistent nuclear activity.

\textit{Global quantities and H\,{\sc i} measurements.} The CIGALE host-galaxy SED fit gives $M_\star=(2.01\pm0.15)\times10^{10}\,M_\odot$, ${\rm SFR}=2.92\pm0.15\,M_\odot\,{\rm yr}^{-1}$, $L_\star=(5.36\pm0.27)\times10^{10}\,L_\odot$, and $L_{\rm dust}\simeq2.75\times10^{10}\,L_\odot$ \citep{Izviekova2026}. The corresponding specific star-formation rate is ${\rm sSFR}=(1.45\pm0.13)\times10^{-10}\,{\rm yr}^{-1}$ [TW]. Its approximate statistical uncertainty was propagated from the stellar-mass and SFR uncertainties without accounting for their covariance.

The nominal stellar mass lies below the adopted MWNeT interval of $(3$--$8)\times10^{10}\,M_\odot$, and its quoted statistical uncertainty does not reach the lower boundary. This discrepancy is retained explicitly in the Stage~IV assessment; agreement in other global properties does not establish full compatibility with the mandatory constraints.

The integrated flux of the adopted GBT profile is $F_{\rm HI}=3.9707\,{\rm Jy\,km\,s^{-1}}$. Using the redshift-corrected relation $M_{\rm HI}=2.356\times10^5D_{\rm L}^2F_{\rm HI}/(1+z)\,M_\odot$, with $D_{\rm L}$ in Mpc and $F_{\rm HI}$ in ${\rm Jy\,km\,s^{-1}}$ \citep{Meyer2017HI}, gives $M_{\rm HI}=6.22\times10^9\,M_\odot$ and $M_{\rm HI}/M_\star\simeq0.309$ [TW].

Including helium with the conventional factor of 1.36 gives $M_{\rm atom}=1.36M_{\rm HI}=8.46\times10^9\,M_\odot$. The corresponding atomic-gas fraction is $f_{\rm atom}=M_{\rm atom}/(M_\star+M_{\rm atom})\simeq0.296$ [TW]. This is not the total cold-gas fraction: without a molecular-gas measurement, it provides only the lower bound $f_{\rm gas}\gtrsim0.296$ for the adopted masses.

The integrated H\,{\sc i} flux-asymmetry parameter, measured about the systemic velocity, is $A_{\rm flux}=\max(F_{\rm low}/F_{\rm high},F_{\rm high}/F_{\rm low})=1.147$ \citep[e.g.][]{Reynolds2020HIasym}. Tests using symmetric profiles and correlated noise, with alternative baseline and integration-window prescriptions, give a symmetric-null probability of $p_{\rm sym}\simeq0.23$ [TW]. The available data therefore do not establish statistically significant global H\,{\sc i} flux asymmetry.

The radio component discussed by \citet{Golay2026} may contain transient-related emission and is not adopted as a homogeneous quiescent host-radio diagnostic. Molecular-gas and halo masses, the radial gas-phase abundance distribution, and spatially resolved H\,{\sc i} structure remain unconstrained. 

This stage is assigned as \textsc{Partial}.

\textit{Panchromatic SED similarity and transient diagnostics} 

The host-galaxy SED is based on aperture-matched photometry from GALEX, Pan-STARRS1, 2MASS, WISE, IRAS, and NVSS \citep{Izviekova2026}. For the Milky Way comparison, we use the baseline host emission: the transient WISE contribution is defined as $\Delta F_W(t)=F_W(t)-F_{W,\rm base}$, and only $F_{W,\rm base}$ enters the host SED. Delayed radio emission is treated separately and excluded from the quantitative comparison.

We compared the CIGALE best-fitting host SED with the intrinsic Milky Way luminosity SED over the common rest-frame interval $0.12$--$5\,\mu{\rm m}$. The logarithmic-slope sequences give Pearson and Spearman coefficients of $r=0.946$ and $r_{\rm S}=0.905$. The median slope difference is $\widetilde{\Delta s}=0.031$, the median absolute difference is ${\rm median}|\Delta s|=0.272$, and the RMS difference is 0.624. Approximately 69\% of the common wavelength-grid points satisfy $|\Delta s|<0.5$ [TW]. These statistics describe agreement between model SED shapes; they do not independently establish similar star-formation histories.

The closest correspondence occurs over $1$--$5\,\mu{\rm m}$, where ${\rm median}|\Delta s|=0.169$ and $r=0.930$. Larger differences over $0.12$--$1\,\mu{\rm m}$ may reflect recent star formation, attenuation, or the stellar-population mix. The quantitative comparison excludes wavelengths beyond $5\,\mu{\rm m}$ and the radio regime. In the current SED treatment, the IRAS measurements enter as upper limits, and the longer-wavelength model continuum is shown only as a model-dependent extension.

Because the homogeneous comparison is restricted to $0.12$--$5\,\mu{\rm m}$, we assign Stage~V=\textsc{Partial}. This SED similarity does not establish a common evolutionary history.

The TDE-like transient WTP14adeqka provides a supplementary probe of recent nuclear activity and the circumnuclear environment at Stage~VII. Its decade-long mid-infrared evolution shows a delayed and broader $W2$ response, colour hysteresis, and declining effective dust temperature \citep{Izviekova2026}. The characteristic delay of $\tau=224\pm45\,{\rm d}$ corresponds to a light-travel scale of $R_{\rm echo}\simeq0.188\pm0.038\,{\rm pc}$. These results constrain circumnuclear reprocessing but do not uniquely determine the dust geometry or establish a Milky Way-like galaxy-wide evolutionary history.

Delayed radio emission provides an additional probe of the transient and its circumnuclear environment \citep{Golay2026}. Its physical origin remains uncertain, and we do not use it as evidence for similarity to the Milky Way's long-term nuclear or assembly history.

These diagnostics constrain recent nuclear activity and circumnuclear reprocessing. They do not provide a complete galaxy-wide evolutionary record: resolved stellar-population histories, radial abundance profiles, outer-disc accretion signatures, the globular-cluster system, and the extended circumgalactic medium remain unconstrained. 

\textit{Rotation-curve similarity.} The calibrated GBT spectrum has a channel spacing of $\Delta v=3.611\,{\rm km\,s^{-1}}$. Following the ADHI/Cosmicflows convention \citep{Courtois2009ADHI,Dupuy2021CF4HI}, we measured the linewidth at 50\% of the mean flux density within the velocity interval containing the central 90\% of the integrated H\,{\sc i} flux, obtaining $W_{m50}=273.92\,{\rm km\,s^{-1}}$. The correction for cosmological stretching and finite spectral resolution, $W^c_{m50}=W_{m50}/(1+z)-2\lambda\Delta v$, with $\lambda=0.25$, gives $W^c_{m50}=267.02\,{\rm km\,s^{-1}}$. 

We applied the turbulence/profile-shape correction $(W_{mx})^2=(W^c_{m50})^2+W_t^2\{1-2\exp[-(W^c_{m50}/W_c)^2]\}-2W^c_{m50}W_t\{1-\exp[-(W^c_{m50}/W_c)^2]\}$, adopting $W_t=9\,{\rm km\,s^{-1}}$ and $W_c=100\,{\rm km\,s^{-1}}$. Correlated-noise Monte Carlo realisations give $W_{mx}=257.94^{+1.96}_{-3.67}\,{\rm km\,s^{-1}}$ [TW].

Using the working photometric inclination from the structural analysis above,  the global linewidth-based rotational-velocity proxy is $V_{\rm rot,W}=W_{mx}/(2\sin i)\simeq149\,{\rm km\,s^{-1}}$ [TW]. This provisional value depends on the adopted disc geometry and requires revision if the inclination changes. It represents an unresolved global velocity scale and cannot be identified with $V_{\max}$, $V_{\rm flat}$, or a spatially resolved circular velocity. We therefore do not use it to establish rotation-curve similarity at Stage~VI.

We independently characterised the integrated profile using the curve-of-growth formalism of \citet{Yu2022HI}. The concentration parameter is $C_V=V_{85}/V_{25}=2.91^{+0.21}_{-0.24}$, and the profile-shape parameter is $K=-0.049\pm0.012$ [TW]. The quoted interval lies below the empirical double-horned threshold of $K=-0.02$, supporting this profile classification. The measured shape is consistent with a rotating atomic-gas disc, although unresolved spectroscopy cannot establish its spatial kinematic regularity \citep{ElBadry2018HI}.

\begin{figure*}[t]
\centering
\includegraphics[width=0.95\textwidth]{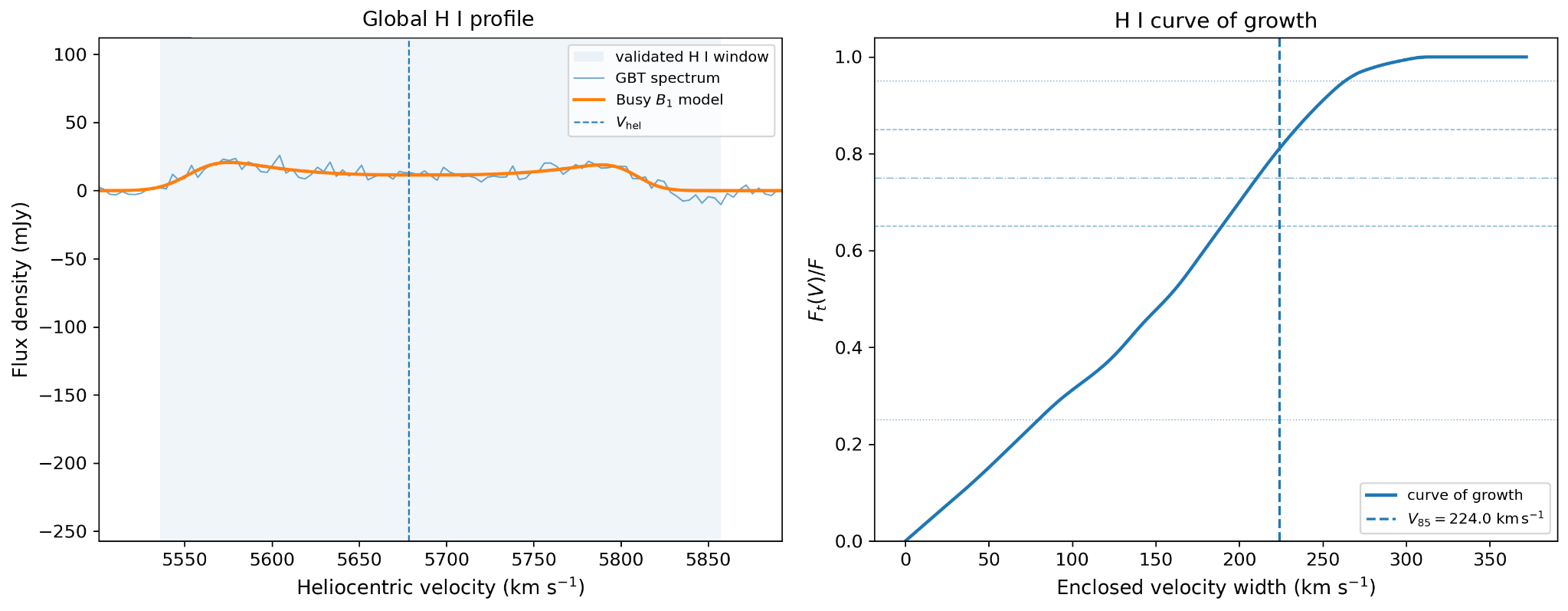}
\caption{UGC~11487 H\,{\sc i} diagnostics. Left: GBT profile, integration window, Busy-function fit, and systemic velocity. Right: normalised curve of growth; the dashed line marks $V_{85}=224.0\,{\rm km\,s^{-1}}$.}
\label{fig:ugc11487_hi_profile}
\end{figure*}

The unresolved H\,{\sc i} spectrum constrains the integrated velocity distribution but not the radial circular-velocity profile $V(R)$, its outer slope, or a disc--halo mass decomposition. A spatially resolved rotation curve suitable for comparison with the Milky Way is therefore unavailable, and Stage~VI remains \textsc{U}.

\textit{Evolutionary fingerprints and fossil tracers.} The infrared and radio evolution of WTP14adeqka constrains recent nuclear activity and circumnuclear reprocessing \citep{Izviekova2026, Golay2026}, but does not establish a Milky-Way-like galaxy-wide evolutionary history. We therefore assign Stage~VII=\textsc{Partial}.

\textit{Integrated MWNeT assessment.} UGC~11487 passes Stages~I--III, showing catalogue-based isolation, a detected stellar bar, and a nuclear state without compelling evidence for a dominant persistent AGN. Its global properties show mixed agreement with the adopted criteria, including a nominal stellar mass below the adopted interval, while its quantitative SED similarity is restricted to $0.12$--$5\,\mu{\rm m}$.  The unresolved H\,{\sc i} profile does not establish rotation-curve similarity, while the transient diagnostics constrain recent nuclear activity without establishing a comparable galaxy-wide assembly history. We retain UGC~11487 as a promising time-domain-selected candidate with explicit qualifications. The principal priorities are resolved gas kinematics, molecular-gas observations, and spatially resolved chemical and stellar-population diagnostics. A dedicated study of UGC~11487 will present the full optical spectral analysis and a more detailed investigation of its nuclear and circumnuclear properties, including the transient and its reprocessing environment.

\section{CGCG~047-026: an early-stage MWNeT candidate requiring follow-up observations}
\label{subsec:cgcg047026}

CGCG~047-026 is included as an illustrative early-stage MWNeT candidate whose archival photometry and spectroscopy permit an assessment of its environment, structure, and nuclear state while its global properties and higher-stage diagnostics remain incompletely constrained. We combine published structural measurements with quantities derived in this work to identify the evidence supporting its candidacy and the priorities for follow-up observations. The adopted measurements and their provenance are compiled in Table~\ref{tab:mwnet_candidates}, with newly derived quantities marked [TW].

\textit{Isolation and environmental context.} CGCG~047-026 was selected from the sample of isolated barred galaxies constructed from the 2MIG parent catalogue. We additionally searched the SDSS spectroscopic data for neighbours satisfying $R_{\rm p}\leq500\,{\rm kpc}$ and $|\Delta V|\leq500\,{\rm km\,s^{-1}}$. At the adopted redshift, the angular-diameter distance is $D_{\rm A}\simeq104.3\,{\rm Mpc}$, corresponding to an angular scale of $1\,{\rm arcsec}\simeq0.506\,{\rm kpc}$. A projected radius of 500~kpc therefore corresponds to approximately $16.5\,{\rm arcmin}$. The adopted velocity interval corresponds approximately to $0.02340\lesssim z\lesssim0.02682$. No SDSS galaxy with a measured redshift satisfying both criteria was found within 500~kpc. The five nearest velocity-selected spectroscopic neighbours lie at projected separations of approximately $1.19$--$1.40\,{\rm Mpc}$, with a fifth-nearest-neighbour distance of $R_{5{\rm NN}}=1.397\,{\rm Mpc}$.

This result supports the catalogue-based isolated status of CGCG~047-026 and motivated its earlier selection as an environmental MWNeT candidate. However, the SDSS spectroscopic census is incomplete for faint and LSB dwarf galaxies; the absence of catalogued neighbours therefore does not rule out a low-mass satellite population (Fig. \ref{fig:stage1_environments}).

We attempted a positional--redshift cross-match with a filament-spine catalogue but could not obtain a reliable nearest-filament distance, \(D_{\mathrm{fil}}\). Both the filament-spine and cosmic-web node distances remain unconstrained. We assign \textsc{Pass} at Stage~I on the basis of catalogue isolation and the absence of a known velocity-selected neighbour within 500~kpc, while retaining the cosmic-web subdiagnostic as \textsc{U}.

\textit{Structural and morphological constraints.} CGCG~047-026 is classified as an SBb galaxy in the decomposition catalogue of \citet{Kruk2018}. The adopted model comprises an exponential disc, a bar, and a central S\'ersic component. Their $r$-band magnitudes, $m_{r,\rm disc}=14.49$, $m_{r,\rm bar}=15.72$, and $m_{r,\rm bulge}=15.68$, give luminosity fractions of $D/T\simeq0.604$, ${\rm Bar}/T\simeq0.194$, and $B/T\simeq0.202$, calculated as $f_j=10^{-0.4m_j}/\sum_k10^{-0.4m_k}$. The catalogue disc effective radius is $r_{e,\rm disc}=38.30$ pixels.  Using the pixel scale of $0.396\,{\rm arcsec\,pixel^{-1}}$ and the exponential-disc relation $r_{e,\rm disc}=1.678h$, we obtain $h\simeq9.04\,{\rm arcsec}\simeq4.57\,{\rm kpc}$ [TW]. This lies near the upper boundary of the adopted MWNeT interval, $2\leq h\leq5\,{\rm kpc}$. The projected disc axis ratio, $q_{\rm disc}=0.84$, implies $i\simeq34^\circ$ for an intrinsic thickness $q_0=0.20$, using $\cos^2i=(q_{\rm disc}^2-q_0^2)/(1-q_0^2)$. This inclination facilitates morphological identification of the bar. The fitted bar has $n_{\rm bar}=0.49$, $q_{\rm bar}=0.39$, and an effective radius of $r_{e,\rm bar}=11.46$ pixels, corresponding to $4.54\,{\rm arcsec}\simeq2.30\,{\rm kpc}$. This effective radius is distinct from the bar semi-major axis; a directly comparable deprojected $a_{\rm bar}$ is not available. The central component has $n_{\rm bulge}=1.48$, $q_{\rm bulge}=0.86$, and $r_{e,\rm bulge}\simeq0.52\,{\rm kpc}$. Its low S\'ersic index is consistent with a pseudobulge-like component, although it does not establish this classification by itself.

The catalogue-based bar identification, compatible disc scale length, and modest central luminosity fraction support \textsc{Pass} at Stage~II. The deprojected bar semi-major axis, bar pattern speed, and secure bulge classification remain unconstrained.

\begin{figure*}[t]
\centering
\includegraphics[width=\textwidth]{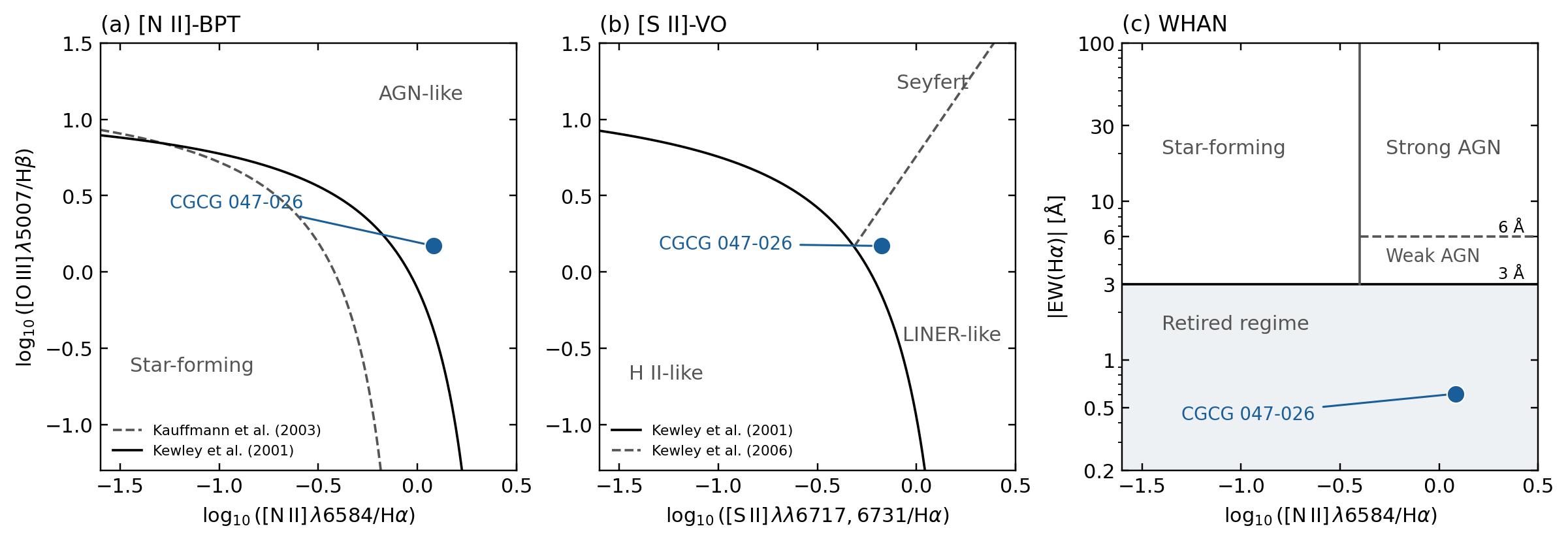}
\caption{Optical emission-line diagnostics of the central SDSS spectrum of CGCG~047-026: (a) [N\,{\sc ii}]-BPT, (b) [S\,{\sc ii}]-VO, and (c) WHAN. Blue points show the adopted catalogue measurements. Demarcation curves follow \citet{Kauffmann2003,Kewley2001,Kewley2006}, and the WHAN boundaries follow \citet{CidFernandes2011}. The [S\,{\sc ii}] diagram indicates LINER-like excitation, while the low rest-frame H$\alpha$ equivalent width, $|{\rm EW}({\rm H}\alpha)|=0.61$~\AA, places the central spectrum in the retired regime. The shaded WHAN region includes retired and passive systems. Measurement uncertainties are not shown because they were not available in the adopted data.}
\label{fig:cgcg_nuclear_diagnostics}
\end{figure*}

\textit{Milky-Way-compatible nuclear state.} The central SDSS spectrum (\texttt{specObjID=2039039150304290816}) gives $\log_{10}([\mathrm{O\,III}]/{\rm H}\beta)=0.170$, $\log_{10}([\mathrm{N\,II}]/{\rm H}\alpha)=0.083$, and $\log_{10}([\mathrm{S\,II}]/{\rm H}\alpha)=-0.176$, where the [S\,{\sc ii}] flux is the sum of the two doublet components [TW]. These ratios indicate LINER-like excitation in the [S\,{\sc ii}] diagnostic diagram (Fig.~\ref{fig:cgcg_nuclear_diagnostics}). However, the low rest-frame H$\alpha$ equivalent width, $|{\rm EW}({\rm H}\alpha)|\simeq0.61$~\AA, places the central spectrum in the retired regime of the WHAN diagram \citep{CidFernandes2011}. The line emission is therefore consistent with ionisation by hot evolved stars and does not by itself establish ongoing black-hole accretion.

The fitted narrow-line velocity dispersions are $\sigma_{\rm Balmer}=109.1\pm18.6\,{\rm km\,s^{-1}}$ and $\sigma_{\rm forbidden}=108.6\pm14.3\,{\rm km\,s^{-1}}$, corresponding to a characteristic ${\rm FWHM}\simeq256\,{\rm km\,s^{-1}}$. No broad Balmer component is identified, precluding a single-epoch broad-line virial black-hole-mass estimate.

The stellar velocity dispersion, $\sigma_\star=154.06\pm3.28\,{\rm km\,s^{-1}}$, gives $M_{\rm BH}\simeq4.4\times10^7\,M_\odot$ [TW] through the relation $\log_{10}(M_{\rm BH}/M_\odot)=8.12+4.24\log_{10}[\sigma_\star/(200\,{\rm km\,s^{-1}})]$ of \citet{Gultekin2009}. The formal propagated uncertainty is approximately 0.04~dex, excluding intrinsic scatter, calibration uncertainties, and possible disc and bar contributions to the aperture measurement.

A complementary estimate follows from the photometric bulge mass. Using the component photometry of \citet{Kruk2018}, the colour--mass-to-light calibration of \citet{Taylor2011}, and $D_{\rm L}=109.6$~Mpc, we obtain $M_{\rm bulge}\simeq1.2\times10^{10}\,M_\odot$ [TW], without a $K$-correction. The \citet{HaringRix2004} relation then gives $M_{\rm BH}\simeq1.5\times10^7\,M_\odot$, while the calibration of \citet{KormendyHo2013} gives approximately $4.2\times10^7\,M_\odot$. These bulge-based estimates require caution because the central component has a low S\'ersic index and is not securely identified as a classical bulge. The resulting spread, $M_{\rm BH}\sim(1.5$--$4.4)\times10^7\,M_\odot$, describes alternative indirect estimates, not a confidence interval.

The measured [O\,{\sc iii}] flux corresponds to $L_{\rm [O\,III]}\simeq7.2\times10^{38}\,{\rm erg\,s^{-1}}$ [TW]. We do not apply an AGN bolometric correction because the line emission may be powered by evolved stars. With no reliable nuclear bolometric or X-ray luminosity, the Eddington ratio remains unconstrained.

The retired central spectrum and absence of a detected broad Balmer component support \textsc{Pass} for the adopted Stage~III nuclear-state criterion. The indirect black-hole masses exceed the reference interval and are retained as a qualification of its similarity to the Milky Way. The complementary diagnostic diagrams are shown in Fig.~\ref{fig:cgcg_nuclear_diagnostics}.

\textit{Global physical constraints.} We estimated the stellar mass from the component photometry of \citet{Kruk2018}, using the colour--mass-to-light relation $\log_{10}(M_\star/L_i)=-0.68+0.70(g-i)$ of \citet{Taylor2011}. The disc, bar, and bulge colours are $(g-i)=1.20$, 1.16, and 1.39, respectively, giving component masses of $M_{\star,\rm disc}\simeq2.8\times10^{10}\,M_\odot$, $M_{\star,\rm bar}\simeq8.1\times10^9\,M_\odot$, and $M_{\star,\rm bulge}\simeq1.2\times10^{10}\,M_\odot$. Their sum, $M_\star\simeq4.8\times10^{10}\,M_\odot$ [TW], lies within the adopted Milky-Way-compatible interval. This estimate remains provisional because systematic uncertainties from stellar-population modelling, attenuation, distance, $K$-corrections, and component decomposition have not been fully quantified.

The SDSS H$\alpha$ measurement samples only the central fibre aperture and does not provide a global star-formation rate. No counterpart was identified in the ALFALFA $\alpha.100$ catalogue within $3\arcmin$ of the optical position. This catalogue non-detection does not establish a quantitative H\,{\sc i} upper limit without an assessment of survey coverage, local sensitivity, and the assumed linewidth. We therefore retain the global SFR, H\,{\sc i} mass, and gas fraction as unconstrained.

No reliable molecular-gas mass, halo mass, resolved circular velocity, radial abundance gradient, or homogeneous dust luminosity is adopted. Although the preliminary stellar mass is compatible with the Milky Way range, the available measurements are insufficient for the combined global assessment. We therefore assign \textsc{U} at Stage~IV.

\textit{Panchromatic SED and rotation-curve similarity.} A homogeneous panchromatic SED has not yet been constructed for CGCG~047-026. The available archival photometry requires aperture matching, consistent foreground-extinction corrections, and a uniform treatment of upper limits before the SED-comparison metrics can be applied. We therefore assign \textsc{U} at Stage~V.

No spatially resolved H\,{\sc i} or optical rotation curve is currently available. The central narrow-line velocity dispersion, $\sigma_{\rm forbidden}\simeq109\,{\rm km\,s^{-1}}$, does not provide a global rotation velocity or constrain the radial dependence $V(R)$. Rotation-curve similarity therefore remains unconstrained, and we assign \textsc{U} at Stage~VI. Spatially resolved H\,{\sc i} observations or extended ionised-gas spectroscopy are required to assess this stage.

\textit{Evolutionary fingerprints and fossil tracers.} The central SDSS spectrum gives $D_n(4000)=2.003\pm0.016$ and $H\delta_{\rm A}=-3.25\pm0.42$~\AA. Together with the weak H$\alpha$ emission, these indices are consistent with an old central stellar population and provide no clear evidence for a strong recent central starburst. This interpretation agrees with the retired spectral classification at Stage~III.

These aperture-limited measurements do not establish the galaxy-wide formation history. Resolved stellar populations, radial chemical abundances, outer accretion signatures, globular clusters, and circumgalactic gas remain insufficiently characterised. We therefore retain the central population indices as supporting evidence but assign \textsc{U} at Stage~VII.

\textit{Integrated MWNeT assessment.} CGCG~047-026 satisfies the adopted environmental, structural, and nuclear-state criteria at Stages~I--III. Its barred-disc structure, compatible disc scale length, and retired central spectrum support its initial candidacy. The preliminary stellar mass is also within the adopted range, while the higher indirect black-hole-mass estimates remain a qualification of its similarity to the Milky Way.

Stages~IV--VII remain unconstrained because the available observations do not support a combined global assessment or tests of SED, rotation-curve, and evolutionary similarity. We therefore classify CGCG~047-026 as an early-stage MWNeT candidate requiring further observations. Priorities include homogeneous panchromatic photometry, resolved gas kinematics and abundance measurements, deep imaging of outer structures, and nuclear X-ray or high-resolution radio observations.

\section{Summary assessment of the candidate galaxies}
\label{sec:summary_assessment}

\begin{table*}[t]
\caption{Summary of the application of the hierarchical Milky Way Near Twin framework to the selected candidate galaxies.}
\label{tab:mwnet_summary}
\centering
\scriptsize
\setlength{\tabcolsep}{2.5pt}
\renewcommand{\arraystretch}{1.15}

\begin{tabular}{
p{0.125\textwidth}
c
c
c
c
c
c
c
p{0.285\textwidth}}
\hline
Galaxy
& I
& II
& III
& IV
& V
& VI
& VII
& Integrated MWNeT assessment \\
\hline

NGC~3521
& \textsc{Pass}
& \textsc{Pass}$^{a}$
& \textsc{Pass}
& \textsc{Pass}
& \textsc{Pass}
& \textsc{Partial}$^{b}$
& \textsc{Pass}$^{c}$
& Strong MWNeT pathfinder; broadly compatible across the hierarchy,
but with an uncertain physical bar identification and incompletely
constrained fossil tracers \\

UGC~11487
& \textsc{Pass}
& \textsc{Pass}
& \textsc{Pass}
& \textsc{Partial}$^{d}$
& \textsc{Partial}$^{j}$
& \textsc{U}$^{e}$
& \textsc{Partial}$^{k}$
& Promising time-domain-selected MWNeT candidate that passes Stages~I--III; its nominal stellar mass lies below the adopted interval, while panchromatic, resolved-kinematic, and galaxy-wide evolutionary assessments remain incomplete \\

NGC~2903
& \textsc{Pass}
& \textsc{Pass}$^{f}$
& \textsc{Pass}
& \textsc{Pass}$^{g}$
& \textsc{Pass}
& \textsc{Pass}
& \textsc{Pass}
& Strong MWNeT pathfinder candidate supported by extensive structural,
kinematic, spectrophotometric, and stellar-population constraints \\

CGCG~047-026
& \textsc{Pass}
& \textsc{Pass}
& \textsc{Pass}$^{h}$
& \textsc{U}$^{i}$
& \textsc{U}
& \textsc{U}
& \textsc{U}
& Promising early-stage candidate that passes Stages~I--III but requires global, gaseous, panchromatic, evolutionary, and resolved-kinematic follow-up \\
\hline
\end{tabular}

\vspace{0.5em}

\begin{minipage}{0.98\textwidth}
\footnotesize
\textit{Notes.}
Stage~I: isolation and cosmic-web context. Stage~II: structural and morphological constraints. Stage~III: Milky Way-compatible nuclear state. Stage~IV: global physical constraints. Stage~V: panchromatic SED similarity. Stage~VI: rotation-curve similarity. Stage~VII: evolutionary fingerprints and fossil tracers. The integrated MWNeT assessment synthesises Stages~I--VII and is not treated as an additional physical stage. \textsc{Pass} denotes consistency with the adopted stage criteria; \textsc{Partial} denotes incomplete, marginal, or definition-sensitive agreement; \textsc{U} denotes insufficient observational constraints.
\medskip
$^{a}$ NGC~3521 satisfies the predefined catalogue-based morphological criterion through its SAB classification. However, the physical identification and deprojection of the putative stellar bar remain uncertain because of the high inclination and the alternative non-barred photometric classification.

$^{b}$ The characteristic rotation amplitude of NGC~3521 is comparable to that of the Milky Way, but its fitted outer rotation curve declines substantially more slowly.

$^{c}$ Several evolutionary diagnostics of NGC~3521 are compatible with the Milky Way regime, including the broken abundance gradient, resolved stellar populations, anomalous H\,{\sc i}, and low-surface-brightness outer structures. However, its globular-cluster system, extended circumgalactic medium, and detailed accretion history remain incompletely constrained.

$^{d}$ The nominal stellar mass of UGC~11487 lies below the adopted MWNeT interval, and its quoted statistical uncertainty does not reach the lower boundary. Stage~IV=\textsc{Partial} denotes mixed agreement among the global diagnostics, not full compatibility with the mandatory constraints. Molecular-gas and halo masses and the radial abundance gradient remain unconstrained.

$^{e}$ The unresolved double-horned H\,{\sc i} profile of UGC~11487 provides a linewidth-based rotational-velocity proxy but cannot replace a spatially resolved rotation curve or be identified directly with $V_{\rm flat}$.

$^{f}$ NGC~2903 has a securely identified strong stellar bar and Milky-Way-compatible disc and bar dimensions. Its unusually small bulge-to-total ratio lies below the adopted Milky Way interval, but this quantity is decomposition-dependent and is treated as a supporting rather than mandatory structural diagnostic.

$^{g}$ Stage~IV compatibility is supported when the literature stellar mass of $3.50\times10^{10}\,M_\odot$ is adopted. Our homogeneous SED estimate, $(1.36\pm0.21)\times10^{10}\,M_\odot$, lies below the adopted interval; the stellar-mass assessment is therefore model-dependent.

$^{h}$ CGCG~047-026 has a retired LINER/LIER spectrum, weak H$\alpha$ emission, and no evidence for a dominant persistent AGN, and therefore satisfies the Milky Way-compatible nuclear-state criterion. Indirect scaling relations give $M_{\rm BH}\sim(1.5$--$4.4)\times10^{7}\,M_\odot$, somewhat above the mass of Sgr~A*. Because this estimate is not dynamical and depends on the adopted scaling relation and bulge decomposition, it is treated as a qualification rather than as a Stage~III rejection criterion.

$^{i}$ The preliminary stellar-mass estimate for CGCG~047-026 is compatible with the adopted Milky Way range, but its global star-formation, gas, halo, chemical, panchromatic, evolutionary, and resolved-kinematic diagnostics remain insufficiently constrained.

$^{j}$ The UV--NIR SED shape of UGC~11487 is quantitatively similar to that of the Milky Way over $0.12$--$5\,\mu{\rm m}$, but the MIR, FIR, and radio regimes are not included in the homogeneous comparison.

$^{k}$ The MIR-selected TDE-like flare, circumnuclear dust response, and delayed radio emission provide evidence for recent transient nuclear activity in UGC~11487. However, its merger history, resolved stellar populations, globular-cluster system, circumgalactic medium, and large-scale fossil structures remain insufficiently constrained.
\end{minipage}
\end{table*}

Each galaxy is assigned one of four assessment classes at every stage:
\begin{itemize}
\item \textsc{Pass}: the available measurements satisfy the adopted Milky Way compatibility criteria for the corresponding stage; 
\item \textsc{Partial}: the galaxy shows broad agreement with the Milky Way, but one or more diagnostics are marginal, definition-sensitive, or not fully consistent; 
\item \textsc{Fail}: at least one mandatory criterion is demonstrably incompatible with the adopted Milky Way range; 
\item \textsc{U} (unconstrained): the available data are insufficient for a reliable assessment of the corresponding stage.
\end{itemize}
The distinction between \textsc{Fail} and \textsc{U} is essential. A candidate lacking resolved H\,{\sc i} kinematics, for example, cannot be regarded as inconsistent with the Milky Way; it remains unconstrained at the corresponding stage. Likewise, a \textsc{Partial} classification does not necessarily imply physical incompatibility, because some structural quantities depend on the adopted decomposition, wavelength, distance, or measurement convention.

The classifications in Table~\ref{tab:mwnet_summary} refer to the combined evidence available at each hierarchical stage. Appendix~A, Table~\ref{tab:mwnet_candidates}, provides the detailed measurements, literature sources, and quantity-specific qualifications for each galaxy.

\section{MWNeTs' Habitable Zone}
\label{sec:mwnets_habitable_zone}

The identification of MWNeTs provides an opportunity to extend the comparison of galaxies beyond their present-day properties towards the astrophysical conditions that may influence the long-term development and persistence of potentially habitable environments. In this context, we introduce the concept of the \textit{MWNeTs' Habitable Zone} as a prospective extension of the Galactic Habitable Zone (GHZ) framework to galaxies with broadly Milky-Way-like structural and evolutionary characteristics.

The GHZ concept was initially developed in the context of Galactic chemical evolution and the metallicity requirements for terrestrial planet formation \citep{Gonzalez2001}. Subsequent models incorporated the time-dependent effects of chemical enrichment, star formation, and supernova activity \citep{Lineweaver2004}. The approach was extended to M31 through chemical evolution models constrained by radial metallicity gradients \citep{Carigi2013}. The inclusion of radial gas flows further demonstrated that the predicted spatial distribution of potentially habitable environments depends on the adopted evolutionary scenario \citep{Spitoni2014}. Complementary dynamical approaches have considered the possible role of spiral-arm encounters and corotation \citep{Marochnik1983}, while numerical models have explored the temporal evolution and spatial distribution of potentially favourable galactic environments \citep{Forgan2016, Vukotic2016}.

Within the MWNeT framework, these approaches can be linked to the evolutionary diagnostics of Stages~V--VII, particularly panchromatic properties, disc dynamics, and fossil tracers of galactic evolution. However, the identification of a galaxy as an MWNeT does not imply that it possesses a GHZ similar to that of the Milky Way. Galactic habitability depends on the spatial and temporal evolution of chemical enrichment, star formation, energetic events, and stellar orbital histories, which cannot be inferred uniquely from present-day galaxy similarity.

NGC~3521 provides a promising case for future GHZ modelling because several complementary evolutionary diagnostics are available. These include the radial oxygen-abundance distribution, with a break at $R_{\rm b,O/H}=4.9\pm0.5$ kpc \citep{Pilyugin2025}, resolved atomic and molecular gas distributions and star formation profiles \citep{Leroy2008}, stellar-population constraints \citep{Coccato2018}, and the observed rotation curve. Together, these data offer an observational basis for exploring the combined effects of chemical enrichment, star formation history, and disc dynamics on the spatial and temporal distribution of potentially habitable environments. However, the available measurements do not yet permit the boundaries of a GHZ to be determined.  We therefore regard the MWNeTs' Habitable Zone as a prospective application of the evolutionary framework.

\section{Conclusions}
\label{sec:conclusions}
We applied the seven-stage Milky Way Near Twins (MWNeT) framework introduced in Paper~I to NGC~3521, NGC~2903, UGC~11487, and CGCG~047-026. These illustrative systems demonstrate how the hierarchy distinguishes present-day compatibility, evidence for evolutionary similarity, and observational incompleteness. They do not constitute a statistically selected sample.

For the first time, these four galaxies are assessed within a common MWNeT hierarchy that combines present-day compatibility with complementary evidence for evolutionary similarity. For UGC~11487, we derive quantitative constraints on its stellar bar, Milky-Way-relative UV--NIR SED shape, and global H\,{\sc i} profile, bringing structural, spectrophotometric, gaseous, and time-domain evidence into a single evolutionary assessment.

NGC~2903 provides a well-constrained benchmark, while NGC~3521 offers extensive evolutionary diagnostics but retains a definition-sensitive structural assessment, particularly concerning its bar. UGC~11487 combines a securely detected stellar bar, quantitative UV--NIR SED similarity, and distinctive transient nuclear tracers; its lower nominal stellar mass and incomplete higher-stage diagnostics qualify its MWNeT interpretation. CGCG~047-026 satisfies the initial environmental, structural, and nuclear-state criteria, but lacks the observations required for a comprehensive higher-stage assessment.

The applications show that complementary diagnostics constrain different aspects of galaxy evolution without uniquely reconstructing an evolutionary pathway. Missing observations must be distinguished from measured incompatibility, and agreement with individual criteria does not establish full MWNeT compatibility. The resulting assessments identify concrete priorities for follow-up observations and provide a basis for extending the framework to systematically selected samples.

Selected candidates also offer external laboratories for investigating the Milky Way's evolution. As a continuation of searches for Earth-like planets and Solar-System analogues, the search for MWNeTs asks whether comparable galactic histories have produced similar long-term environments for planetary systems. Quantitative assessment of this possibility requires spatially resolved evolutionary diagnostics and dedicated models of galactic habitability.

\begin{acknowledgements}
We thank Dr. Walter Golay for providing the reduced MMT/Binospec data products. The observations were obtained at the MMT Observatory, a joint facility of the University of Arizona and the Smithsonian Institution, using Binospec under programme SAO-13-24A (PI: Daichi Hiramatsu).

This study was supported by the National Research Foundation of Ukraine (Project No. 2023.03/0188). This research has made extensive use of the SAO/NASA Astrophysics Data System. 
\end{acknowledgements}

\bibliographystyle{aa} 
\bibliography{MWNeTs} 

@article{ArgudoFernandez2015,
  author  = {Argudo-Fern{\'a}ndez, M. and Verley, S. and Bergond, G. and Sulentic, J. W. and Sabater, J. and Fern{\'a}ndez Lorenzo, M. and Espada, D. and Leon, S. and S{\'a}nchez-Exp{\'o}sito, S. and Lisenfeld, U. and Verdes-Montenegro, L.},
  title   = {Catalogues of Isolated Galaxies, Isolated Pairs, and Isolated Triplets in the Local Universe},
  journal = {Astronomy \& Astrophysics},
  year    = {2015},
  volume  = {578},
  eid     = {A110},
  pages   = {A110},
  doi     = {10.1051/0004-6361/201526016}
}

@MISC{Buta2015Cat,
  author       = {{Buta}, R.~J. and {Sheth}, K. and {Athanassoula}, E. and {Bosma}, A. and {Knapen}, J.~H. and {Laurikainen}, E. and {Salo}, H. and {Elmegreen}, D. and {Ho}, L.~C. and {Zaritsky}, D. and {Courtois}, H. and {Hinz}, J.~L. and {Mu{\~n}oz-Mateos}, J.-C. and {Kim}, T. and {Regan}, M.~W. and {Gadotti}, D.~A. and {Gil de Paz}, A. and {Laine}, J. and {Men{\'e}ndez-Delmestre}, K. and {Comer{\'o}n}, S. and {Erroz-Ferrer}, S. and {Seibert}, M. and {Mizusawa}, T. and {Holwerda}, B. and {Madore}, B.~F.},
  title        = {{VizieR Online Data Catalog: S4G galaxy morphologies in the CVRHS system}},
  howpublished = {VizieR Online Data Catalog, J/ApJS/217/32},
  year         = {2015},
  eid          = {J/ApJS/217/32},
  doi          = {10.26093/cds/vizier.22170032},
  adsurl       = {https://ui.adsabs.harvard.edu/abs/2015yCat..22170032B}
}

@article{Carigi2013,
  author  = {Carigi, L. and Garc{\'i}a-Rojas, J.
             and Meneses-Goytia, S.},
  title   = {Chemical Evolution and the Galactic
             Habitable Zone of M31},
  journal = {Revista Mexicana de Astronom{\'i}a
             y Astrof{\'i}sica},
  year    = {2013},
  volume  = {49},
  pages   = {253--273},
  eprint  = {1208.4198},
  archivePrefix = {arXiv}
}

@article{Carlsten2022,
  author        = {Carlsten, Scott G. and Greene, Jenny E. and
                   Beaton, Rachael L. and Danieli, Shany and
                   Greco, Johnny P.},
  title         = {The Exploration of Local VolumE Satellites
                   ({ELVES}) Survey: A Nearly Volume-limited Sample
                   of Nearby Dwarf Satellite Systems},
  journal       = {The Astrophysical Journal},
  year          = {2022},
  volume        = {933},
  number        = {1},
  eid           = {47},
  pages         = {47},
  doi           = {10.3847/1538-4357/ac6fd7},
  eprint        = {2203.00014},
  archivePrefix = {arXiv},
  primaryClass  = {astro-ph.GA}
}

@article{CidFernandes2011,
  author  = {{Cid Fernandes}, R. and Stasi{\'n}ska, G. and
             Mateus, A. and {Vale Asari}, N.},
  title   = {A comprehensive classification of galaxies in the
             Sloan Digital Sky Survey: how to tell true from fake AGN?},
  journal = {MNRAS},
  year    = {2011},
  volume  = {413},
  number  = {3},
  pages   = {1687--1699},
  doi     = {10.1111/j.1365-2966.2011.18244.x}
}

@article{Golay2026,
  author        = {Golay, Walter W. and Berger, Edo and Cendes, Yvette
                   and Masterson, Megan and Polisensky, Emil
                   and Mutel, Robert L. and Blanchard, Peter K.
                   and Kumar, Harsh and Margutti, Raffaella
                   and Drout, Maria and Panagiotou, Christos
                   and De, Kishalay and Kara, Erin},
  title         = {Radio Emission from the Infrared Tidal Disruption
                   Event WTP14adeqka: A Directly Resolved
                   Nonrelativistic Outflow},
  journal       = {The Astrophysical Journal},
  year          = {2026},
  volume        = {1007},
  number        = {1},
  eid           = {88},
  pages         = {88},
  doi           = {10.3847/1538-4357/ae8536},
  eprint        = {2508.16756},
  archivePrefix = {arXiv},
  primaryClass  = {astro-ph.HE}
}

@ARTICLE{Das2003,
       author = {{Das}, Mousumi and {Teuben}, Peter J. and {Vogel}, Stuart N. and {Regan}, Michael W. and {Sheth}, Kartik and {Harris}, Andrew I. and {Jefferys}, William H.},
        title = "{Central Mass Concentration and Bar Dissolution in Nearby Spiral Galaxies}",
      journal = {\apj},
         year = 2003,
        month = jan,
       volume = {582},
       number = {1},
        pages = {190-195},
          doi = {10.1086/344480},
archivePrefix = {arXiv},
       eprint = {astro-ph/0208467},
 primaryClass = {astro-ph},
       adsurl = {https://ui.adsabs.harvard.edu/abs/2003ApJ...582..190D}
}

@article{Davis2014,
  author        = {Davis, Benjamin L. and Berrier, Joel C. and
                   Johns, Lucas and Shields, Douglas W. and
                   Hartley, Matthew T. and Kennefick, Daniel and
                   Kennefick, Julia and Seigar, Marc S. and
                   Lacy, Claud H. S.},
  title         = {The Black Hole Mass Function Derived from
                   Local Spiral Galaxies},
  journal       = {The Astrophysical Journal},
  year          = {2014},
  volume        = {789},
  number        = {2},
  eid           = {124},
  pages         = {124},
  doi           = {10.1088/0004-637X/789/2/124},
  eprint        = {1405.5876},
  archivePrefix = {arXiv},
  primaryClass  = {astro-ph.CO}
}

@BOOK{deVaucouleurs1991RC3,
  author    = {de Vaucouleurs, Gerard and de Vaucouleurs, Antoinette
               and Corwin, Harold G. Jr. and Buta, Ronald J.
               and Paturel, Georges and Fouque, Pascal},
  title     = {Third Reference Catalogue of Bright Galaxies},
  year      = {1991},
  publisher = {Springer-Verlag},
  address   = {New York},
  doi       = {10.1007/978-1-4757-4363-0}
}

@ARTICLE{Dobrycheva2026,
       author = {{Dobrycheva}, D.~V. and {Vavilova}, I.~B. and {Kompaniiets}, O.~V. and {Khramtsov}, V. and {Vasylenko}, M. Yu. and {Hetmantsev}, O.~O. and {Melnyk}, O.~V. and {Karachentseva}, V.~E.},
        title = "{Machine learning technique for morphological classification of galaxies from SDSS. IV. Visual inspection vs CNN for merging, irregular, edge-on, barred, ringed, and with dust lanes galaxies at 0.02<z<0.1}",
      journal = {arXiv e-prints},
         year = 2026,
        month = apr,
          eid = {arXiv:2604.24471},
        pages = {arXiv:2604.24471},
          doi = {10.48550/arXiv.2604.24471},
archivePrefix = {arXiv},
       eprint = {2604.24471},
 primaryClass = {astro-ph.GA},
       adsurl = {https://ui.adsabs.harvard.edu/abs/2026arXiv260424471D}
}

@article{Forgan2016,
  author  = {Forgan, D. and Dayal, P. and Cockell, C. S.
             and Libeskind, N.},
  title   = {Evaluating galactic habitability using high-resolution
             cosmological simulations of galaxy formation},
  journal = {International Journal of Astrobiology},
  year    = {2017},
  volume  = {16},
  number  = {1},
  pages   = {60--73},
  doi     = {10.1017/S1473550415000518}
}

@article{Gonzalez2001,
  author  = {Gonzalez, G. and Brownlee, D. and Ward, P.},
  title   = {The Galactic Habitable Zone: Galactic Chemical Evolution},
  journal = {Icarus},
  year    = {2001},
  volume  = {152},
  number  = {1},
  pages   = {185--200},
  doi     = {10.1006/icar.2001.6617}
}

@ARTICLE{Gultekin2009,
  author  = {{G{\"u}ltekin}, K. and {Richstone}, D.~O. and {Gebhardt}, K. and
             {Lauer}, T.~R. and {Tremaine}, S. and {Aller}, M.~C. and
             {Bender}, R. and {Dressler}, A. and {Faber}, S.~M. and
             {Filippenko}, A.~V. and {Green}, R. and {Ho}, L.~C. and
             {Kormendy}, J. and {Magorrian}, J. and {Pinkney}, J. and
             {Siopis}, C.},
  title   = {The $M$--$\sigma$ and $M$--$L$ Relations in Galactic Bulges, and Determinations of Their Intrinsic Scatter},
  journal = {The Astrophysical Journal},
  year    = {2009},
  volume  = {698},
  number  = {1},
  pages   = {198--221},
  doi     = {10.1088/0004-637X/698/1/198},
  eprint  = {0903.4897},
  archivePrefix = {arXiv},
  primaryClass  = {astro-ph.GA}
}

@ARTICLE{HaringRix2004,
  author  = {{H{\"a}ring}, N. and {Rix}, H.-W.},
  title   = {On the Black Hole Mass--Bulge Mass Relation},
  journal = {\apjl},
  year    = {2004},
  volume  = {604},
  number  = {2},
  pages   = {L89--L92},
  doi     = {10.1086/383567},
  eprint  = {astro-ph/0402376},
  archivePrefix = {arXiv}
}

@article{HernandezToledo2010,
  author  = {Hern{\'a}ndez-Toledo, H. M. and V{\'a}zquez-Mata, J. A. and Mart{\'i}nez-V{\'a}zquez, L. A. and Choi, Y.-Y. and Park, C.},
  title   = {The UNAM-KIAS Catalog of Isolated Galaxies},
  journal = {The Astronomical Journal},
  year    = {2010},
  volume  = {139},
  number  = {6},
  pages   = {2525--2541},
  doi     = {10.1088/0004-6256/139/6/2525}
}

@article{Ho2008,
  author  = {Ho, L. C.},
  title   = {Nuclear Activity in Nearby Galaxies},
  journal = {ARA\&A},
  volume  = {46},
  pages   = {475--539},
  year    = {2008},
  doi     = {10.1146/annurev.astro.45.051806.110546}
}

@article{Inman2026,
  author  = {Inman, Christopher J. and Popescu, Cristina C. and Murphy, David},
  title   = {The galaxies' energy balance problem solved},
  journal = {Monthly Notices of the Royal Astronomical Society},
  year    = {2026},
  volume  = {551},
  number  = {1},
  pages   = {1--33},
  doi     = {10.1093/mnras/stag1342},
  eprint  = {2608.14023},
  archivePrefix = {arXiv},
  primaryClass  = {astro-ph.GA}
}

@ARTICLE{Izviekova2026,
       author = {{Izviekova}, I.~O. and {Kompaniiets}, O.~V. and {Vavilova}, I.~B.},
        title = "{Revealing a multi-zone circumnuclear dust response in UGC\raisebox{-0.5ex}\textasciitilde11487 through phase-resolved WISE W1--W2 diagnostics}",
      journal = {arXiv e-prints},
         year = 2026,
        month = jul,
          eid = {arXiv:2607.17815},
        pages = {arXiv:2607.17815},
          doi = {10.48550/arXiv.2607.17815},
archivePrefix = {arXiv},
       eprint = {2607.17815},
 primaryClass = {astro-ph.GA},
       adsurl = {https://ui.adsabs.harvard.edu/abs/2026arXiv260717815I}
}

@article{Karachentsev2022NGC3115,
  author        = {Karachentsev, I. D. and Makarova, L. N. and
                   Anand, Gagandeep S. and Tully, R. Brent},
  title         = {Around the Spindle Galaxy: The Dark Halo Mass
                   of NGC 3115},
  journal       = {The Astronomical Journal},
  year          = {2022},
  volume        = {163},
  number        = {5},
  eid           = {234},
  pages         = {234},
  doi           = {10.3847/1538-3881/ac5ab5},
  eprint        = {2203.01700},
  archivePrefix = {arXiv},
  primaryClass  = {astro-ph.GA}
}

@article{Karachentseva1973,
  author  = {Karachentseva, V. E.},
  title   = {Catalogue of Isolated Galaxies},
  journal = {Soobshcheniya Spetsial'noj Astrofizicheskoj Observatorii},
  year    = {1973},
  volume  = {8},
  pages   = {3--49},
  adsurl  = {https://ui.adsabs.harvard.edu/abs/1973SoSAO...8....3K}
}

@article{Karachentseva2010,
  author  = {Karachentseva, V. E. and Mitronova, S. N. and Melnyk, O. V. and Karachentsev, I. D.},
  title   = {Catalog of Isolated Galaxies Selected from the 2MASS Survey},
  journal = {Astrophysical Bulletin},
  year    = {2010},
  volume  = {65},
  number  = {1},
  pages   = {1--12},
  doi     = {10.1134/S1990341310010019},
  eprint  = {1005.3191},
  archivePrefix = {arXiv},
  primaryClass  = {astro-ph.CO}
}

@ARTICLE{Karachentseva2011Companions,
  author  = {Karachentseva, V. E. and Karachentsev, I. D. and Melnyk, O. V.},
  title   = {Faint Companions of Isolated 2MIG Galaxies},
  journal = {Astrophysical Bulletin},
  year    = {2011},
  volume  = {66},
  number  = {4},
  pages   = {389--406},
  doi     = {10.1134/S1990341311040018},
  eprint  = {1112.2821},
  archivePrefix = {arXiv},
  primaryClass  = {astro-ph.CO}
}

@article{Karachentsev2011LOG,
  author  = {Karachentsev, I. D. and Makarov, D. I. and Karachentseva, V. E. and Melnyk, O. V.},
  title   = {Catalog of Nearby Isolated Galaxies in the Volume $z<0.01$},
  journal = {Astrophysical Bulletin},
  year    = {2011},
  volume  = {66},
  number  = {1},
  pages   = {1--27},
  doi     = {10.1134/S1990341311010019}
}

@article{Kauffmann2003,
  author  = {Kauffmann, G. and Heckman, T. M. and Tremonti, C. and
             Brinchmann, J. and Charlot, S. and White, S. D. M. and
             Ridgway, S. E. and Brinkmann, J. and Fukugita, M. and
             Hall, P. B. and Ivezi{\'c}, {\v Z}. and Richards, G. T. and
             Schneider, D. P.},
  title   = {The Host Galaxies of Active Galactic Nuclei},
  journal = {MNRAS},
  year    = {2003},
  volume  = {346},
  pages   = {1055--1077},
  doi     = {10.1111/j.1365-2966.2003.07154.x}
}

@article{Kewley2001,
  author  = {Kewley, L. J. and Dopita, M. A. and Sutherland, R. S. and
             Heisler, C. A. and Trevena, J.},
  title   = {Theoretical Modeling of Starburst Galaxies},
  journal = {ApJ},
  year    = {2001},
  volume  = {556},
  pages   = {121--140},
  doi     = {10.1086/321545}
}

@article{Kewley2006,
  author  = {Kewley, L. J. and Groves, B. and Kauffmann, G. and
             Heckman, T.},
  title   = {The Host Galaxies and Classification of Active Galactic Nuclei},
  journal = {MNRAS},
  year    = {2006},
  volume  = {372},
  pages   = {961--976},
  doi     = {10.1111/j.1365-2966.2006.10859.x}
}

@ARTICLE{Kompaniiets2025,
       author = {{Kompaniiets}, O.~V. and {Kukhar}, O.~M. and {Vavilova}, I.~B. and {Dobrycheva}, D.~V. and {Fedorov}, P.~N. and {Dmytrenko}, A.~M. and {Khramtsov}, V.~P. and {Sergijenko}, O.~M. and {Vasylenko}, A.~A.},
        title = {Milky Way galaxy-analogs and isolated galaxies with bars: environmental density in the Local Volume},
      journal = {Space Sci. Technol.},
         year = 2025,
        volume = {31},
        number = {6},
        pages = {134--148},
        doi = {10.15407/knit2025.06.134}
}

@ARTICLE{Kompaniiets2026,
   author = {{Kompaniiets}, O.~V. and {Vavilova}, I.~B. and {Vasylkivskyi}, Y.~V. and {Konovalenko}, O.~O. and {Pastoven}, O.~S. and {Izviekova}, I.~O. and {Junais} and {Dmytrenko}, A.~M. and {Dobrycheva}, D.~V. and {Fedorov}, P.~N. and {Khramtsov}, V.~P. and {Sergijenko}, O. and {Vasylenko}, A.~A.},
        title = "{NGC 3521 as the Milky Way near-twin: UV─radio-decameter spectral energy distribution}",
      journal = {\aap},
         year = 2026,
        month = may,
       volume = {710},
          eid = {A11},
        pages = {A11},
          doi = {10.1051/0004-6361/202558597},
archivePrefix = {arXiv},
       eprint = {2512.14664},
 primaryClass = {astro-ph.GA},
       adsurl = {https://ui.adsabs.harvard.edu/abs/2026A&A...710A..11K}
}

@article{KormendyHo2013,
  author = {Kormendy, J. and Ho, L. C.},
  title = {Coevolution (Or Not) of Supermassive Black Holes and Host Galaxies},
  journal = {ARA\&A},
  volume = {51},
  pages = {511--653},
  year = {2013}
}

@article{Kourkchi2017,
  author        = {Kourkchi, Ehsan and Tully, R. Brent},
  title         = {Galaxy Groups Within 3500 km s$^{-1}$},
  journal       = {The Astrophysical Journal},
  year          = {2017},
  volume        = {843},
  number        = {1},
  eid           = {16},
  pages         = {16},
  doi           = {10.3847/1538-4357/aa76db},
  eprint        = {1705.08068},
  archivePrefix = {arXiv},
  primaryClass  = {astro-ph.GA}
}

@article{Kruk2018,
  author  = {Kruk, Sandor J. and Lintott, Chris J. and
             Bamford, Steven P. and Masters, Karen L. and
             Simmons, Brooke D. and H{\"a}u{\ss}ler, Boris and
             Cardamone, Carolin N. and Hart, Ross E. and
             Kelvin, Lee and Schawinski, Kevin and
             Smethurst, Rebecca J. and Vika, Marina},
  title   = {Galaxy Zoo: secular evolution of barred galaxies
             from structural decomposition of multiband images},
  journal = {Monthly Notices of the Royal Astronomical Society},
  year    = {2018},
  volume  = {473},
  number  = {4},
  pages   = {4731--4753},
  doi     = {10.1093/mnras/stx2605},
  eprint  = {1710.00093},
  archivePrefix = {arXiv},
  primaryClass  = {astro-ph.GA}
}

@article{Lineweaver2004,
  author  = {Lineweaver, C. H. and Fenner, Y. and Gibson, B. K.},
  title   = {The Galactic Habitable Zone and the Age Distribution
             of Complex Life in the Milky Way},
  journal = {Science},
  year    = {2004},
  volume  = {303},
  number  = {5654},
  pages   = {59--62},
  doi     = {10.1126/science.1092322}
}

@article{Marochnik1983,
  author  = {Marochnik, L. S.},
  title   = {On the Origin of the Solar System and the
             Exceptional Position of the Sun in the Galaxy},
  journal = {Astrophysics and Space Science},
  year    = {1983},
  volume  = {89},
  pages   = {61--75},
  doi     = {10.1007/BF01008385}
}

@ARTICLE{MartinezDelgado2010,
  author  = {{Mart{\'i}nez-Delgado}, D. and {Gabany}, R.~J. and {Crawford}, K. and {Zibetti}, S. and {Majewski}, S.~R. and {Rix}, H.-W. and {Fliri}, J. and {Carballo-Bello}, J.~A. and {Bardalez-Gagliuffi}, D.~C. and {Pe{\~n}arrubia}, J. and {Chonis}, T.~S. and {Madore}, B. and {Trujillo}, I. and {Schirmer}, M. and {McDavid}, D.~A.},
  title   = {Stellar Tidal Streams in Spiral Galaxies of the Local Volume: A Pilot Survey with Modest Aperture Telescopes},
  journal = {\aj},
  year    = {2010},
  volume  = {140},
  number  = {4},
  pages   = {962--967},
  doi     = {10.1088/0004-6256/140/4/962},
  eprint  = {1003.4860},
  archivePrefix = {arXiv},
  primaryClass  = {astro-ph.CO}
}

@ARTICLE{MartnezDelgado2025,
       author = {{Mart{\'\i}nez-Delgado}, David and {Stein}, Michael and {Sakowska}, Joanna D. and {Maurice Weigelt}, M. and {Rom{\'a}n}, Javier and {Donatiello}, Giuseppe and {Roca-F{\`a}brega}, Santi and {Schirmer}, Mischa and {Grebel}, Eva K. and {Saifollahi}, Teymoor and {Kanipe}, Jeff and {G{\'o}mez-Flechoso}, M. Angeles and {Akhlaghi}, Mohammad and {Javanmardi}, Behnam and {Wu}, Gang and {Eskandarlou}, Sepideh and {Bomans}, Dominik J. and {Henkel}, Cristian and {Block}, Adam and {Hanson}, Mark and {Schedler}, Johannes and {Teuwen}, Karel and {GaBany}, R. Jay and {Iba{\~n}ez Perez}, Alvaro and {Crawford}, Ken and {Promper}, Wolfgang and {Jimenez}, Manuel and {Farr{\`a}s-Aloy}, S{\'\i}lvia and {Mir{\'o}-Carretero}, Juan},
        title = "{Stellar tidal streams around nearby spiral galaxies with deep imaging from amateur telescopes}",
      journal = {\aap},
         year = 2025,
        month = sep,
       volume = {701},
          eid = {A182},
        pages = {A182},
          doi = {10.1051/0004-6361/202554980},
archivePrefix = {arXiv},
       eprint = {2504.02071},
 primaryClass = {astro-ph.GA},
       adsurl = {https://ui.adsabs.harvard.edu/abs/2025A&A...701A.182M}
}

@ARTICLE{Masterson2024,
       author = {{Masterson}, Megan and {De}, Kishalay and {Panagiotou}, Christos and {Kara}, Erin and {Arcavi}, Iair and {Eilers}, Anna-Christina and {Frostig}, Danielle and {Gezari}, Suvi and {Grotova}, Iuliia and {Liu}, Zhu and {Malyali}, Adam and {Meisner}, Aaron M. and {Merloni}, Andrea and {Newsome}, Megan and {Rau}, Arne and {Simcoe}, Robert A. and {van Velzen}, Sjoert},
        title = "{A New Population of Mid-infrared-selected Tidal Disruption Events: Implications for Tidal Disruption Event Rates and Host Galaxy Properties}",
      journal = {\apj},
         year = 2024,
        month = feb,
       volume = {961},
       number = {2},
          eid = {211},
        pages = {211},
          doi = {10.3847/1538-4357/ad18bb},
archivePrefix = {arXiv},
       eprint = {2401.01403},
 primaryClass = {astro-ph.HE},
       adsurl = {https://ui.adsabs.harvard.edu/abs/2024ApJ...961..211M}
}

@ARTICLE{McGaugh2016,
       author = {{McGaugh}, Stacy S.},
        title = "{The Surface Density Profile of the Galactic Disk from the Terminal Velocity Curve}",
      journal = {\apj},
         year = 2016,
        month = jan,
       volume = {816},
       number = {1},
          eid = {42},
        pages = {42},
          doi = {10.3847/0004-637X/816/1/42},
archivePrefix = {arXiv},
       eprint = {1511.06387},
 primaryClass = {astro-ph.GA},
       adsurl = {https://ui.adsabs.harvard.edu/abs/2016ApJ...816...42M}
}

@ARTICLE{Muller2025,
  author        = {{M{\"u}ller}, O. and {Jerjen}, H. and {Taibi}, S. and {Heesters}, N. and {Crosby}, E. and {Pawlowski}, M.~S.},
  title         = {New dwarf galaxy candidates in the M106, NGC~3521, and UGCA~127 groups with the Hyper Suprime-Cam},
  journal       = {arXiv e-prints},
  year          = {2025},
  month         = apr,
  eid           = {arXiv:2504.11608},
  eprint        = {2504.11608},
  archivePrefix = {arXiv},
  primaryClass  = {astro-ph.GA},
  doi           = {10.48550/arXiv.2504.11608}
}

@ARTICLE{Natale2022,
       author = {{Natale}, Giovanni and {Popescu}, Cristina C. and {Rushton}, Mark and {Yang}, Ruizhi and {Thirlwall}, Jordan J. and {Pricopi}, Dumitru},
        title = "{A radiation transfer model for the Milky Way: II. The global properties and large-scale structure}",
      journal = {MNRAS},
         year = 2022,
        month = jan,
       volume = {509},
       number = {2},
        pages = {2339-2361},
          doi = {10.1093/mnras/stab2771},
archivePrefix = {arXiv},
       eprint = {2111.02180},
 primaryClass = {astro-ph.GA},
       adsurl = {https://ui.adsabs.harvard.edu/abs/2022MNRAS.509.2339N}
}

@ARTICLE{Ou2024,
       author = {{Ou}, Xiaowei and {Eilers}, Anna-Christina and {Necib}, Lina and {Frebel}, Anna},
        title = "{The dark matter profile of the Milky Way inferred from its circular velocity curve}",
      journal = {\mnras},
         year = 2024,
        month = feb,
       volume = {528},
       number = {1},
        pages = {693-710},
          doi = {10.1093/mnras/stae034},
archivePrefix = {arXiv},
       eprint = {2303.12838},
 primaryClass = {astro-ph.GA},
       adsurl = {https://ui.adsabs.harvard.edu/abs/2024MNRAS.528..693O}
}

@article{Pastoven2024,
author = {Pastoven, O. S. and Kompaniiets, O. and Vavilova, I. and others},
title = {NGC 3521 as the Milky Way analogue: spectral energy distribution from UV to radio and photometric variability},
journal = {Space Sci. \& Technol.},
volume = {30},
number = {6},
pages = {67--83},
year = {2024},
doi = {10.15407/knit2024.06.067}
}

@ARTICLE{Paturel2003,
  author  = {{Paturel}, G. and {Petit}, C. and {Prugniel}, P. and {Theureau}, G. and {Rousseau}, J. and {Brouty}, M. and {Dubois}, P. and {Cambr{\'e}sy}, L.},
  title   = {HYPERLEDA. I. Identification and designation of galaxies},
  journal = {\aap},
  year    = {2003},
  volume  = {412},
  pages   = {45--55},
  doi     = {10.1051/0004-6361:20031411}
}

@article{Pilyugin2023,
  author        = {Pilyugin, L. S. and Tautvai{\v{s}}ien{\.e}, G.
                   and Lara-L{\'o}pez, M. A.},
  title         = {Searching for Milky Way Twins:
                   Radial Abundance Distribution as a Strict Criterion},
  journal       = {Astronomy \& Astrophysics},
  year          = {2023},
  volume        = {676},
  eid           = {A57},
  pages         = {A57},
  doi           = {10.1051/0004-6361/202346503},
  eprint        = {2306.09854},
  archivePrefix = {arXiv},
  primaryClass  = {astro-ph.GA},
  adsurl        = {https://ui.adsabs.harvard.edu/abs/2023A%26A...676A..57P}
}

@ARTICLE{Pilyugin2025,
       author = {{Pilyugin}, L.~S. and {Lara-L{\'o}pez}, M.~A. and {Tautvai{\v{s}}ien{\.{e}}}, G. and {Zinchenko}, I.~A. and {Gardu{\~n}o}, L.~E. and {De Rossi}, M.~E. and {Zaragoza-Cardiel}, J. and {Dib}, S. and {Val{\'e}}, G.},
        title = "{Metal-THINGS: The Milky Way twin candidate NGC 3521}",
      journal = {\aap},
         year = 2025,
        month = feb,
       volume = {694},
          eid = {A113},
        pages = {A113},
          doi = {10.1051/0004-6361/202452605},
archivePrefix = {arXiv},
       eprint = {2501.07443},
 primaryClass = {astro-ph.GA},
       adsurl = {https://ui.adsabs.harvard.edu/abs/2025A&A...694A.113P}
}

@ARTICLE{Regan2006,
       author = {{Regan}, Michael W. and {Thornley}, Michele D. and {Vogel}, Stuart N. and {Sheth}, Kartik and {Draine}, Bruce T. and {Hollenbach}, David J. and {Meyer}, Martin and {Dale}, Daniel A. and {Engelbracht}, Charles W. and {Kennicutt}, Robert C. and {Armus}, Lee and {Buckalew}, Brent and {Calzetti}, Daniela and {Gordon}, Karl D. and {Helou}, George and {Leitherer}, Claus and {Malhotra}, Sangeeta and {Murphy}, Eric and {Rieke}, George H. and {Rieke}, Marcia J. and {Smith}, J.~D.},
        title = "{The Radial Distribution of the Interstellar Medium in Disk Galaxies: Evidence for Secular Evolution}",
      journal = {\apj},
         year = 2006,
        month = dec,
       volume = {652},
       number = {2},
        pages = {1112-1121},
          doi = {10.1086/505382},
       adsurl = {https://ui.adsabs.harvard.edu/abs/2006ApJ...652.1112R}
}

@article{ReinesVolonteri2015,
  author = {Reines, Amy E. and Volonteri, Marta},
  title = {Relations between Central Black Hole Mass and Total Galaxy
           Stellar Mass in the Local Universe},
  journal = {The Astrophysical Journal},
  year = {2015},
  volume = {813},
  number = {2},
  eid = {82},
  pages = {82},
  doi = {10.1088/0004-637X/813/2/82},
  eprint = {1508.06274},
  archivePrefix = {arXiv},
  primaryClass = {astro-ph.GA},
  adsurl = {https://ui.adsabs.harvard.edu/abs/2015ApJ...813...82R}
}

@ARTICLE{Salak2019,
       author = {{Salak}, Dragan and {Noma}, Yuto and {Sorai}, Kazuo and {Miyamoto}, Yusuke and {Kuno}, Nario and {Pettitt}, Alex R. and {Kaneko}, Hiroyuki and {Tanaka}, Takahiro and {Yasuda}, Atsushi and {Kita}, Shoichiro and {Yajima}, Yoshiyuki and {Shibata}, Shugo and {Nakai}, Naomasa and {Seta}, Masumichi and {Muraoka}, Kazuyuki and {Kuroda}, Mayu and {Nakanishi}, Hiroyuki and {Takeuchi}, Tsutomu T. and {Yoda}, Moe and {Morokuma-Matsui}, Kana and {Watanabe}, Yoshimasa and {Matsumoto}, Naoko and {Oi}, Nagisa and {Pan}, Hsi-An and {Kajikawa}, Ayumi and {Yashima}, Yu and {Komatsuzaki}, Ryusei},
        title = "{CO Multi-line Imaging of Nearby Galaxies (COMING). VII. Fourier decomposition of molecular gas velocity fields and bar pattern speed}",
      journal = {\pasj},
         year = 2019,
        month = dec,
       volume = {71},
          eid = {S16},
        pages = {S16},
          doi = {10.1093/pasj/psz004},
archivePrefix = {arXiv},
       eprint = {1901.00640},
 primaryClass = {astro-ph.GA},
       adsurl = {https://ui.adsabs.harvard.edu/abs/2019PASJ...71S..16S}
}

@ARTICLE{Salo2015,
       author = {{Salo}, Heikki and {Laurikainen}, Eija and {Laine}, Jarkko and {Comer{\'o}n}, Sebastien and {Gadotti}, Dimitri A. and {Buta}, Ron and {Sheth}, Kartik and {Zaritsky}, Dennis and {Ho}, Luis and {Knapen}, Johan and {Athanassoula}, E. and {Bosma}, Albert and {Laine}, Seppo and {Cisternas}, Mauricio and {Kim}, Taehyun and {Mu{\~n}oz-Mateos}, Juan Carlos and {Regan}, Michael and {Hinz}, Joannah L. and {Gil de Paz}, Armando and {Menendez-Delmestre}, Karin and {Mizusawa}, Trisha and {Erroz-Ferrer}, Santiago and {Meidt}, Sharon E. and {Querejeta}, Miguel},
        title = "{The Spitzer Survey of Stellar Structure in Galaxies (S$^{4}$G): Multi-component Decomposition Strategies and Data Release}",
      journal = {\apjs},
         year = 2015,
        month = jul,
       volume = {219},
       number = {1},
          eid = {4},
        pages = {4},
          doi = {10.1088/0067-0049/219/1/4},
archivePrefix = {arXiv},
       eprint = {1503.06550},
 primaryClass = {astro-ph.GA},
       adsurl = {https://ui.adsabs.harvard.edu/abs/2015ApJS..219....4S}
}

@article{Spitoni2014,
  author  = {Spitoni, E. and Matteucci, F. and Sozzetti, A.},
  title   = {The galactic habitable zone of the Milky Way and M31
             from chemical evolution models with gas radial flows},
  journal = {MNRAS},
  year    = {2014},
  volume  = {440},
  number  = {3},
  pages   = {2588--2598},
  doi     = {10.1093/mnras/stu484}
}

@ARTICLE{Tabatabaei2017,
author = {{Tabatabaei}, F.~S. and and others},
title = {High-resolution radio continuum survey of M33 and the radio--infrared correlation},
journal = {ApJ},
year = 2017,
volume = {836},
pages = {185},
doi = {10.3847/1538-4357/836/2/185}
}

@article{TeboulMetzger2023,
  author  = {Teboul, Odelia and Metzger, Brian D.},
  title   = {A Unified Theory of Jetted Tidal Disruption Events:
             From Promptly Escaping Relativistic to Delayed
             Transrelativistic Jets},
  journal = {The Astrophysical Journal Letters},
  year    = {2023},
  volume  = {957},
  number  = {1},
  pages   = {L9},
  doi     = {10.3847/2041-8213/ad0037},
  eprint  = {2308.05161},
  archivePrefix = {arXiv},
  primaryClass  = {astro-ph.HE}
}

@ARTICLE{Taylor2011,
  author  = {{Taylor}, E.~N. and {Hopkins}, A.~M. and {Baldry}, I.~K. and {Brown}, M.~J.~I. and {Driver}, S.~P. and {Kelvin}, L.~S. and {Hill}, D.~T. and {Robotham}, A.~S.~G. and {Bland-Hawthorn}, J. and {Jones}, D.~H. and {Sharp}, R.~G. and {Thomas}, D. and {Liske}, J. and {Loveday}, J. and {Norberg}, P. and {Peacock}, J.~A. and {Bamford}, S.~P. and {Brough}, S. and {Colless}, M. and {Cameron}, E. and {Conselice}, C.~J. and {Croom}, S.~M. and {Frenk}, C.~S. and {Gunawardhana}, M. and {Kuijken}, K. and {Nichol}, R.~C. and {Parkinson}, H.~R. and {Phillipps}, S. and {Pimbblet}, K.~A. and {Popescu}, C.~C. and {Prescott}, M. and {Sutherland}, W.~J. and {Tuffs}, R.~J. and {van Kampen}, E. and {Wijesinghe}, D.},
  title   = {Galaxy And Mass Assembly (GAMA): stellar mass estimates},
  journal = {\mnras},
  year    = {2011},
  volume  = {418},
  number  = {3},
  pages   = {1587--1620},
  doi     = {10.1111/j.1365-2966.2011.19536.x},
  eprint  = {1108.0635},
  archivePrefix = {arXiv},
  primaryClass  = {astro-ph.CO}
}

@article{Tully1977,
author = {Tully, R. B. and Fisher, J. R.},
title = {A New Method of Determining Distances to Galaxies},
journal = {A\&A},
volume = {54},
pages = {661--673},
year = {1977}
}

@ARTICLE{Vavilova2024,
       author = {{Vavilova}, I.~B. and {Fedorov}, P.~M. and {Dobrycheva}, D.~V. and {Sergijenko}, O. and {Vasylenko}, A.~A. and {Dmytrenko}, A.~M. and {Khramtsov}, V.~P. and {Kompaniiets}, O.~V.},
        title = "{An advanced approach to the definition of the ``Milky Way galaxies-analogues''}",
      journal = {Space Sci. Technol.},
         year = 2024,
        month = jan,
       volume = {30},
       number = {4},
        pages = {81},
          doi = {10.15407/knit2024.04.081},
archivePrefix = {arXiv},
       eprint = {2410.04411},
 primaryClass = {astro-ph.GA},
       adsurl = {https://ui.adsabs.harvard.edu/abs/2024KosNT..30d..81V}
}

@ARTICLE{Vavilova2026,
       author = {{Vavilova}, I.~B. and {Dmytrenko}, A.~M. and {Dobrycheva}, D.~V. and {Fedorov}, P.~N. and {Izviekova}, I.~O. and {Khramtsov}, V.~P. and {Kompaniiets}, O.~V. and {Kukhar}, O.~N. and {Pastoven}, O.~S. and {Sergijenko}, O. and {Vasylenko}, A.~A.},
        title = "{Milky Way Near Twins (MWNeTs). I. A Hierarchical Framework for Identifying the Evolutionary Counterparts of the Milky Way}",
      journal = {arXiv e-prints},
         year = 2026,
        month = jul,
          eid = {arXiv:2607.08336},
        pages = {arXiv:2607.08336},
          doi = {10.48550/arXiv.2607.08336},
archivePrefix = {arXiv},
       eprint = {2607.08336},
 primaryClass = {astro-ph.GA},
       adsurl = {https://ui.adsabs.harvard.edu/abs/2026arXiv260708336V}
}

@article{Veilleux1987,
  author  = {Veilleux, S. and Osterbrock, D. E.},
  title   = {Spectral Classification of Emission-Line Galaxies},
  journal = {ApJS},
  year    = {1987},
  volume  = {63},
  pages   = {295--310},
  doi     = {10.1086/191166}
}

@article{VerdesMontenegro2005,
  author  = {Verdes-Montenegro, L. and Sulentic, J. and Lisenfeld, U. and Leon, S. and Espada, D. and Garcia, E. and Sabater, J. and Verley, S.},
  title   = {The AMIGA project. I. Optical characterization of the CIG},
  journal = {Astronomy \& Astrophysics},
  year    = {2005},
  volume  = {436},
  pages   = {443--455},
  doi     = {10.1051/0004-6361:20042280}
}

@article{Vukotic2016,
  author  = {Vukoti{\'c}, B. and Steinhauser, D.
             and Martinez-Aviles, G.
             and {\'C}irkovi{\'c}, M. M.
             and Micic, M. and Schindler, S.},
  title   = {{'Grandeur in this view of life'}:
             N-body simulation models of the Galactic
             habitable zone},
  journal = {Monthly Notices of the Royal Astronomical Society},
  year    = {2016},
  volume  = {459},
  number  = {4},
  pages   = {3512--3524},
  doi     = {10.1093/mnras/stw829}
}

@article{Yuan2014,
  author  = {Yuan, F. and Narayan, R.},
  title   = {Hot Accretion Flows Around Black Holes},
  journal = {ARA\&A},
  volume  = {52},
  pages   = {529--588},
  year    = {2014},
  doi     = {10.1146/annurev-astro-082812-141003}
}

@article{Zhang2009,
  author        = {Zhang, Wei Ming and Soria, Roberto and
                   Zhang, Shuang Nan and Swartz, Douglas A. and
                   Liu, JiFeng},
  title         = {A Census of X-Ray Nuclear Activity in Nearby Galaxies},
  journal       = {The Astrophysical Journal},
  year          = {2009},
  volume        = {699},
  number        = {1},
  pages         = {281--297},
  doi           = {10.1088/0004-637X/699/1/281},
  eprint        = {0904.1091},
  archivePrefix = {arXiv},
  primaryClass  = {astro-ph.HE}
}

@ARTICLE{Zhou2012,
       author = {{Zhou}, Zhi-Min and {Cao}, Chen and {Wu}, Hong},
        title = "{Star formation properties in barred galaxies (SFB) II. NGC 2903 and NGC 7080}",
      journal = {Research in Astronomy and Astrophysics},
         year = 2012,
        month = mar,
       volume = {12},
       number = {3},
        pages = {235-248},
          doi = {10.1088/1674-4527/12/3/001},
archivePrefix = {arXiv},
       eprint = {1111.3411},
 primaryClass = {astro-ph.CO},
       adsurl = {https://ui.adsabs.harvard.edu/abs/2012RAA....12..235Z}
}

@ARTICLE{Coccato2018,
  author = {Coccato, Lodovico and Fabricius, Maximilian H. and
            Saglia, Roberto P. and Bender, Ralf and Erwin, Peter and
            Drory, Niv and Morelli, Lorenzo},
  title = {Spectroscopic decomposition of NGC 3521: unveiling the properties of the bulge and disc},
  journal = {Monthly Notices of the Royal Astronomical Society},
  year = {2018},
  volume = {477},
  number = {2},
  pages = {1958--1969},
  doi = {10.1093/mnras/sty705},
  eprint = {1803.03653},
  archivePrefix = {arXiv},
  primaryClass = {astro-ph.GA}
}

@ARTICLE{Muraoka2016,
  author = {Muraoka, Kazuyuki and Sorai, Kazuo and Kuno, Nario and
            Nakai, Naomasa and Nakanishi, Hiroyuki and Takeda, Miho and
            Yanagitani, Kazuki and Kaneko, Hiroyuki and Miyamoto, Yusuke and
            Kishida, Nozomi and Hatakeyama, Takuya and Umei, Michiko and
            Tanaka, Takahiro and Tomiyasu, Yuto and Saita, Chey and
            Ueno, Saeko and Matsumoto, Naoko and Salak, Dragan and
            Morokuma-Matsui, Kana},
  title = {{CO} Multi-line Imaging of Nearby Galaxies ({COMING}). I.
           Physical properties of molecular gas in the barred spiral galaxy
           {NGC 2903}},
  journal = {Publications of the Astronomical Society of Japan},
  year = {2016},
  volume = {68},
  number = {5},
  pages = {89},
  doi = {10.1093/pasj/psw080},
  eprint = {1607.08023},
  archivePrefix = {arXiv}
}

@ARTICLE{Grisdale2017,
  author = {Grisdale, Kearn and Agertz, Oscar and Romeo, Alessandro B. and
            Renaud, Florent and Read, Justin I.},
  title = {The impact of stellar feedback on the density and velocity
           structure of the interstellar medium},
  journal = {Monthly Notices of the Royal Astronomical Society},
  year = {2017},
  volume = {466},
  number = {1},
  pages = {1093--1110},
  doi = {10.1093/mnras/stw3133}
}

@ARTICLE{Carrillo2020,
  author = {Carrillo, Andreia and Jogee, Shardha and Drory, Niv and
            Kaplan, Kyle F. and Blanc, Guillermo and Weinzirl, Tim and
            Song, Mimi and Luo, Rongxin},
  title = {The VIRUS-P Exploration of Nearby Galaxies (VENGA):
           the stellar populations and assembly of NGC 2903's bulge,
           bar, and outer disc},
  journal = {Monthly Notices of the Royal Astronomical Society},
  year = {2020},
  volume = {493},
  number = {3},
  pages = {4094--4106},
  doi = {10.1093/mnras/staa397}
}

@ARTICLE{Irwin2009,
  author = {Irwin, Judith A. and Hoffman, G. Lyle and Spekkens, Kristine and
            Haynes, Martha P. and Giovanelli, Riccardo and Linder, Suzanne M. and
            Catinella, Barbara and Momjian, Emmanuel and Koribalski, B{\"a}rbel S. and
            Davies, Jonathan and Brinks, Elias and de Blok, W. J. G. and
            Putman, Mary E. and van Driel, Wim},
  title = {$\Lambda$CDM Satellites and H I Companions---The Arecibo
           ALFA Survey of NGC 2903},
  journal = {The Astrophysical Journal},
  year = {2009},
  volume = {692},
  number = {2},
  pages = {1447--1463},
  doi = {10.1088/0004-637X/692/2/1447}
}

@ARTICLE{Leroy2008,
  author = {Leroy, Adam K. and Walter, Fabian and Brinks, Elias and
            Bigiel, Frank and de Blok, W. J. G. and Madore, Barry and
            Thornley, M. D.},
  title = {The Star Formation Efficiency in Nearby Galaxies:
           Measuring Where Gas Forms Stars Effectively},
  journal = {The Astronomical Journal},
  year = {2008},
  volume = {136},
  number = {6},
  pages = {2782--2845},
  doi = {10.1088/0004-6256/136/6/2782},
  eprint = {0810.2556},
  archivePrefix = {arXiv}
}

@article{Karachentsev2015LeoSpur,
  author  = {Karachentsev, I. D. and Tully, R. B. and Makarova, L. N. and Makarov, D. I. and Rizzi, L.},
  title   = {Peculiar Velocities of Galaxies in the Leo Spur},
  journal = {The Astrophysical Journal},
  year    = {2015},
  volume  = {805},
  number  = {2},
  pages   = {144},
  doi     = {10.1088/0004-637X/805/2/144},
  eprint  = {1503.06027},
  archivePrefix = {arXiv},
  primaryClass  = {astro-ph.GA}
}

@article{She2017ChandraCatalog,
  author  = {She, Rui and Ho, Luis C. and Feng, Hua},
  title   = {Chandra Survey of Nearby Galaxies: The Catalog},
  journal = {The Astrophysical Journal},
  year    = {2017},
  volume  = {835},
  number  = {2},
  pages   = {223},
  doi     = {10.3847/1538-4357/835/2/223},
  eprint  = {1612.08507},
  archivePrefix = {arXiv},
  primaryClass  = {astro-ph.GA}
}

@ARTICLE{Zeilinger2001,
  author  = {Zeilinger, W. W. and Vega Beltr{\'a}n, J. C. and
             Rozas, M. and Beckman, J. E. and Pizzella, A. and
             Corsini, E. M. and Bertola, F.},
  title   = {{NGC 3521}: Stellar Counter-Rotation Induced by a Bar Component},
  journal = {Astrophysics and Space Science},
  year    = {2001},
  volume  = {276},
  number  = {2-4},
  pages   = {643--650},
  doi     = {10.1023/A:1017548101623}
}

@ARTICLE{Zurita2021,
  author  = {Zurita, A. and Florido, E. and Bresolin, F. and
             P{\'e}rez-Montero, E. and P{\'e}rez, I.},
  title   = {Bar Effect on Gas-Phase Abundance Gradients. I.
             Data Sample and Chemical Abundances},
  journal = {Monthly Notices of the Royal Astronomical Society},
  year    = {2021},
  volume  = {500},
  number  = {2},
  pages   = {2359--2379},
  doi     = {10.1093/mnras/staa2246},
  eprint  = {2007.12289},
  archivePrefix = {arXiv},
  primaryClass  = {astro-ph.GA}
}

@ARTICLE{HessmanZiebart2011,
  author  = {Hessman, F. V. and Ziebart, M.},
  title   = {The Bosma Effect Revisited. I. H\,{\sc i} and
             Stellar Disc Scaling Models},
  journal = {Astronomy \& Astrophysics},
  year    = {2011},
  volume  = {532},
  pages   = {A121},
  doi     = {10.1051/0004-6361/201117199},
  eprint  = {1106.5621},
  archivePrefix = {arXiv},
  primaryClass  = {astro-ph.CO}
}

@ARTICLE{Elson2014,
  author  = {Elson, E. C.},
  title   = {An H\,{\sc i} Study of {NGC 3521} -- A Galaxy
             with a Slow-Rotating Halo},
  journal = {Monthly Notices of the Royal Astronomical Society},
  year    = {2014},
  volume  = {437},
  number  = {4},
  pages   = {3736--3749},
  doi     = {10.1093/mnras/stt2182},
  eprint  = {1312.2399},
  archivePrefix = {arXiv},
  primaryClass  = {astro-ph.GA}
}

@ARTICLE{Chyzy2018,
  author  = {Chy{\.z}y, K. T. and Jurusik, W. and Piotrowska, J. and
             Nikiel-Wroczy{\'n}ski, B. and Heesen, V. and Vacca, V. and
             Nowak, N. and Paladino, R. and Surma, P. and Sridhar, S. S. and
             Heald, G. and Beck, R. and Conway, J. and Sendlinger, K. and
             Cury{\l}o, M. and Mulcahy, D. and Broderick, J. W. and
             Hardcastle, M. J. and Callingham, J. R. and G{\"u}rkan, G. and
             Iacobelli, M. and R{\"o}ttgering, H. J. A. and Adebahr, B. and
             Shulevski, A. and Dettmar, R.-J. and Breton, R. P. and
             Clarke, A. O. and Farnes, J. S. and Orr{\'u}, E. and
             Pandey, V. N. and Pandey-Pommier, M. and Pizzo, R. and
             Riseley, C. J. and Rowlinson, A. and Scaife, A. M. M. and
             Stewart, A. J. and van der Horst, A. J. and van Weeren, R. J.},
  title   = {LOFAR MSSS: Flattening low-frequency radio continuum spectra
             of nearby galaxies},
  journal = {Astronomy \& Astrophysics},
  year    = {2018},
  volume  = {619},
  pages   = {A36},
  doi     = {10.1051/0004-6361/201833133},
  eprint  = {1808.10374},
  archivePrefix = {arXiv},
  primaryClass  = {astro-ph.GA}
}

@ARTICLE{SylosLabini2024,
  author  = {Sylos Labini, Francesco and De Marzo, Giordano and
             Straccamore, Matteo and Comer{\'o}n, S{\'e}bastien},
  title   = {The Tully--Fisher relation and the Bosma effect},
  journal = {Monthly Notices of the Royal Astronomical Society},
  year    = {2024},
  volume  = {527},
  number  = {2},
  pages   = {2697--2717},
  doi     = {10.1093/mnras/stad3278},
  eprint  = {2310.15673},
  archivePrefix = {arXiv},
  primaryClass  = {astro-ph.GA}
}

@ARTICLE{Yukita2012,
  author  = {Yukita, Mihoko and Swartz, Douglas A. and
             Tennant, Allyn F. and Soria, Roberto and
             Irwin, Jimmy A.},
  title   = {Hot Diffuse Emission in the Nuclear Starburst Region
             of NGC 2903},
  journal = {The Astrophysical Journal},
  year    = {2012},
  volume  = {758},
  number  = {2},
  pages   = {105},
  doi     = {10.1088/0004-637X/758/2/105},
  eprint  = {1209.5443},
  archivePrefix = {arXiv},
  primaryClass  = {astro-ph.CO}
}

@article{Fedorov2026,
  author = {Fedorov, P. N. and Dmytrenko, A. M. and Akhmetov, V. S. and Velichko, A. B. and Zheleznyak, A. P. and Kompaniiets, O. V.},
  title = {Local Jeans estimates of the Galaxy's circular velocity, based on RGB stars from the Gaia catalogue},
  journal = {MNRAS},
  volume = {550},
  number = {4},
  pages = {stag1310},
  year = {2026},
  doi = {10.1093/mnras/stag1310}
}

@article{Lelli2016,
  author = {Lelli, Federico and McGaugh, Stacy S. and Schombert, James M.},
  title = {SPARC: Mass Models for 175 Disk Galaxies with Spitzer Photometry and Accurate Rotation Curves},
  journal = {AJ},
  volume = {152},
  number = {6},
  pages = {157},
  year = {2016},
  doi = {10.3847/0004-6256/152/6/157},
  eprint = {1606.09251},
  archivePrefix = {arXiv},
  primaryClass = {astro-ph.GA}
}

@article{deBlok2008,
  author = {de Blok, W. J. G. and Walter, Fabian and Brinks, Elias and Trachternach, Carsten and Oh, Se-Heon and Kennicutt, Robert C.},
  title = {High-resolution rotation curves and galaxy mass models from THINGS},
  journal = {AJ},
  volume = {136},
  number = {6},
  pages = {2648--2719},
  year = {2008},
  doi = {10.1088/0004-6256/136/6/2648},
  eprint = {0810.2100},
  archivePrefix = {arXiv},
  primaryClass = {astro-ph}
}

@ARTICLE{Dey2019Legacy,
  author = {{Dey}, Arjun and {Schlegel}, David J. and {Lang}, Dustin and
            {Blum}, Robert and {Burleigh}, Kaylan and {Fan}, Xiaohui and
            {Findlay}, Joseph R. and {Finkbeiner}, Doug and {Herrera}, David and
            {Juneau}, St{\'e}phanie and {Landriau}, Martin and {Levi}, Michael and
            {McGreer}, Ian and {Meisner}, Aaron and {Myers}, Adam D. and
            {Moustakas}, John and {Nugent}, Peter and {Patej}, Anna and
            {Schlafly}, Edward F. and {Walker}, Alistair R. and others},
  title = "{Overview of the DESI Legacy Imaging Surveys}",
  journal = {\aj},
  year = 2019,
  volume = {157},
  number = {5},
  eid = {168},
  pages = {168},
  doi = {10.3847/1538-3881/ab089d},
  archivePrefix = {arXiv},
  eprint = {1804.08657},
  primaryClass = {astro-ph.IM}
}

@ARTICLE{Hall2012,
  author = {{Hall}, Melanie and {Courteau}, St{\'e}phane and
            {Dutton}, Aaron A. and {McDonald}, Michael and {Zhu}, Yucong},
  title = "{An investigation of Sloan Digital Sky Survey imaging data
            and multiband scaling relations of spiral galaxies}",
  journal = {\mnras},
  year = 2012,
  volume = {425},
  number = {4},
  pages = {2741--2765},
  doi = {10.1111/j.1365-2966.2012.21290.x}
}

@ARTICLE{PhipsonSmyth2010,
  author = {{Phipson}, Belinda and {Smyth}, Gordon K.},
  title = "{Permutation P-values Should Never Be Zero:
            Calculating Exact P-values When Permutations Are Randomly Drawn}",
  journal = {Statistical Applications in Genetics and Molecular Biology},
  year = 2010,
  volume = {9},
  number = {1},
  eid = {39},
  pages = {Article 39},
  doi = {10.2202/1544-6115.1585}
}

@ARTICLE{Courtois2009ADHI,
  author = {{Courtois}, H{\'e}l{\`e}ne M. and {Tully}, R. Brent and
            {Fisher}, J. Richard and {Bonhomme}, Nicolas and
            {Zavodny}, Maximilian and {Barnes}, Austin},
  title = "{The Extragalactic Distance Database:
            All Digital H I Profile Catalog}",
  journal = {\aj},
  year = 2009,
  volume = {138},
  number = {6},
  pages = {1938--1956},
  doi = {10.1088/0004-6256/138/6/1938},
  archivePrefix = {arXiv},
  eprint = {0902.3670},
  primaryClass = {astro-ph.CO}
}

@ARTICLE{Dupuy2021CF4HI,
  author = {{Dupuy}, A. and {Courtois}, H. M. and {Guinet}, D. and
            {Tully}, R. B. and {Kourkchi}, E.},
  title = "{Toward Cosmicflows-4: The H I data catalog}",
  journal = {\aap},
  year = 2021,
  volume = {646},
  eid = {A113},
  pages = {A113},
  doi = {10.1051/0004-6361/202039025},
  archivePrefix = {arXiv},
  eprint = {2101.01515},
  primaryClass = {astro-ph.GA}
}

@ARTICLE{Meyer2017HI,
  author = {{Meyer}, Martin and {Robotham}, Aaron and
            {Obreschkow}, Danail and {Westmeier}, Tobias and
            {Duffy}, Alan R. and {Staveley-Smith}, Lister},
  title = "{Tracing H I Beyond the Local Universe}",
  journal = {Publications of the Astronomical Society of Australia},
  year = 2017,
  volume = {34},
  eid = {e052},
  pages = {e052},
  doi = {10.1017/pasa.2017.31}
}

@ARTICLE{Reynolds2020HIasym,
  author = {{Reynolds}, T. N. and {Westmeier}, T. and
            {Staveley-Smith}, L. and {Chauhan}, G. and {Lagos}, C. D. P.},
  title = "{H I asymmetries in LVHIS, VIVA, and HALOGAS galaxies}",
  journal = {\mnras},
  year = 2020,
  volume = {493},
  number = {4},
  pages = {5089--5106},
  doi = {10.1093/mnras/staa597}
}

@ARTICLE{Yu2022HI,
  author = {{Yu}, Niankun and {Ho}, Luis C. and
            {Wang}, Jing and {Li}, Hangyuan},
  title = "{Statistical Analysis of H I Profile Asymmetry and Shape
            for Nearby Galaxies}",
  journal = {\apjs},
  year = 2022,
  volume = {261},
  number = {2},
  eid = {21},
  pages = {21},
  doi = {10.3847/1538-4365/ac626b},
  archivePrefix = {arXiv},
  eprint = {2203.13404},
  primaryClass = {astro-ph.GA}
}

@ARTICLE{ElBadry2018HI,
  author = {{El-Badry}, Kareem and {Bradford}, Jeremy and
            {Quataert}, Eliot and {Geha}, Marla and
            {Boylan-Kolchin}, Michael and {Weisz}, Daniel R. and
            {Wetzel}, Andrew and {Hopkins}, Philip F. and
            {Chan}, T. K. and {Fitts}, Alex and
            {Kere{\v{s}}}, Du{\v{s}}an and
            {Faucher-Gigu{\`e}re}, Claude-Andr{\'e}},
  title = "{Gas kinematics in FIRE simulated galaxies compared to
            spatially unresolved H I observations}",
  journal = {\mnras},
  year = 2018,
  volume = {477},
  number = {2},
  pages = {1536--1548},
  doi = {10.1093/mnras/sty730},
  archivePrefix = {arXiv},
  eprint = {1801.03933},
  primaryClass = {astro-ph.GA}
}

\appendix

\clearpage
\onecolumn

\section{Application of the hierarchical framework to select MWNeT candidates}
\label{app:application_MWNet}

\begingroup
\footnotesize
\setlength{\LTpre}{0pt}
\setlength{\LTpost}{0.5em}
\setlength{\tabcolsep}{3pt}
\renewcommand{\arraystretch}{1.12}

\begin{longtable}{
@{}
>{\raggedright\arraybackslash}p{0.13\textwidth}
>{\raggedright\arraybackslash}p{0.207\textwidth}
>{\raggedright\arraybackslash}p{0.207\textwidth}
>{\raggedright\arraybackslash}p{0.207\textwidth}
>{\raggedright\arraybackslash}p{0.207\textwidth}
@{}}

\caption{Application of the hierarchical MWNeT framework to selected candidate galaxies}
\label{tab:mwnet_candidates}\\
\endfirsthead

\multicolumn{5}{c}{\tablename~\thetable\ -- continued}\\
\hline
& \textbf{NGC~3521} & \textbf{UGC~11487} & \textbf{NGC~2903} & \textbf{CGCG 047-026} \\
\hline
\endhead

\hline
\multicolumn{5}{r}{\footnotesize Continued on the next page}\\
\endfoot

\hline
\endlastfoot

\multicolumn{5}{c}{%
\textbf{Stage I. Low-density Environment and Cosmic-Web Context
($ {\rm MWA}\cap{\rm Isolated} $)}}\\
\hline
& \textbf{NGC~3521} & \textbf{UGC~11487} & \textbf{NGC~2903} & \textbf{CGCG 047-026} \\
& $166.45,\,-0.036$ & $297.35,\,63.50$   & $143.04,\,21.50$  & $215.48,\,7.88$ \\
& $0.0026$          & $0.019$            & $0.0018$          & $0.0251$ \\
\hline

Isolation criterion
& Isolated (KIG~461) [21]
& Isolated (2MIG 2701) [22]
& Isolated (KIG~347) [21]
& Isolated (2MIG 1971) [22]\\

Local environment
& $\Sigma_{\rm Vor}\simeq0.7$ and $\Sigma_{5{\rm NN}}\simeq0.4\,{\rm gal\,Mpc^{-2}}$; a low-to-intermediate projected density broadly comparable to the environmental regime adopted for the Milky Way [14]
& No galaxy with a catalogued redshift was found within   $R_{\rm p}<500$~kpc and   $|\Delta V|<500~{\rm km\,s^{-1}}$ in our control search; the faint-satellite census remains incomplete [TW]
& Nearby non-interacting system with no known comparable-luminosity neighbour; deep optical and H\,{\sc i} surveys reveal only low-mass companions [15,36]
& No spectroscopically confirmed SDSS neighbour within $R_{\rm p}<500$ kpc and $|\Delta V|\leq500~{\rm km\,s^{-1}}$; $R_{5{\rm NN}}=1.397$ Mpc [14,TW]\\

Cosmic-web location
& Located in the diffuse Leo Spur within the low-density Local-Volume environment [9,14,35]
& No robust filament, wall, or void assignment is currently available; the filament-spine and node distances remain unconstrained [TW]
& Located in the nearer strand of the Leo Spur [27]
& Low-density environment selected through the $5-NN$ diagnostic; direct filament, wall, or void assignment remains unconstrained [14] \\

Satellite environment
& No known comparable-mass companion; 11 dwarf satellites with SBF- or redshift-based membership are reported in the available catalogues. Additional LSB dwarf candidates have been identified in projection, but their physical association with NGC~3521 remains to be confirmed [36,37]
& No known comparable-mass companion under the 2MIG isolation criterion; the faint and low-surface-brightness satellite population remains unconstrained [22,TW]
& No comparable-mass companion; seven dwarf satellites are flagged as confirmed in the ELVES catalogue, while the deep H\,{\sc i} survey identifies only low-mass gaseous companions [15,36]
& No spectroscopically confirmed satellite within $R_{\rm p}<500$ kpc; the faint-satellite population remains unconstrained because of spectroscopic incompleteness [TW] \\
\hline

\multicolumn{5}{c}{%
\textbf{Stage II. Structural and Morphological Constraints
($ {\rm MWA}\cap{\rm Isolated}\cap{\rm Barred} $)}}\\
\hline
& \textbf{NGC~3521} & \textbf{UGC~11487} & \textbf{NGC~2903} & \textbf{CGCG~047-026} \\
\hline

Morphological type
& SAB(rs)bc [1], but SA(r'l,r)bc in the CVRHS/S$^4$G classification [2]
& SBb [45]; barred disc with a winding inner $m=2$ component [TW]
& SAB(rs)bc [1]; (R')SB(rs,nr)b [2]
& SBb [47] \\

Barred spiral structure
& Multi-arm structure; weak or bar-like inner component [3]; an almost end-on bar was proposed to explain the central LOSVD [7], but the photometric classification, subsequent spectroscopy, and high inclination do not provide an unambiguous bar identification [2,24]
& Photometrically resolved stellar bar embedded in a winding inner spiral structure; secure photometric detection; $p_{\rm bar}\simeq2.0\times10^{-4}$ from empirical null simulations [TW]
& Prominent stellar bar, particularly strong in the NIR range, with associated grand-design spiral structure [16]
& Bar confirmed by an SDSS $ugriz$ disc+bar+bulge decomposition [47] \\

Disc scale length
& $R_{\rm d}=1.7\pm0.2$ kpc from a single-exponential Spitzer $3.6\,\mu$m disc fit [32]; $R_{\rm d}=2.9$ kpc for the adopted distance $D=10.7$ Mpc [4]; a multi-component radiative-transfer decomposition gives $R_{\rm d}\simeq3.3$--$3.4$ kpc for the main stellar disc [34]
& $h=2.78^{+0.15}_{-0.28}$ kpc from the outer exponential-disc fit [TW]; $i\simeq59.6^\circ$ [TW]; working photometric inclination[TW]
& $R_{\rm d}=2.94$ kpc [5]
& $h\simeq4.57$ kpc, derived from $r_{e,\rm disc}=38.30$ SDSS pixels [TW]; near the upper boundary of the adopted MWNeT range \\

Bulge-to-total ratio
& $B/T=0.104$ from the S$^4$G $3.6\,\mu$m structural decomposition [6]
& Unconstrained; projected disc axis ratio $q_{\rm disc}=0.526^{+0.028}_{-0.023}$ [TW]' $a_{\rm bar}/h=1.71^{+0.27}_{-0.16}$ [TW]
&  $B/T=0.0453$ [5]; Partial assessment -- below the adopted Milky Way interval and consistent with a very small central spheroidal component
& $B/T\simeq0.202$ in the $r$ band [47,TW] \\

Bar structure
& Suspected weak, nearly end-on bar; its intrinsic geometry and deprojected ellipticity remain unconstrained because of the high inclination [7,24]
& Truncated Ferrers-type stellar bar embedded in a winding inner $m=2$ component [TW]; $q_{\rm bar}\simeq0.43$ and $e_{\rm bar}\simeq0.57$, deprojected; secure photometric detection; $p_{\rm bar}\simeq2.0\times10^{-4}$ [TW]
& Strong NIR bar with a disky or pseudobulge-like central component [16]
& $Bar/T\simeq0.194$, $n_{\rm bar}=0.49$, and $(b/a)_{\rm bar}=0.39$ [47,TW] \\

Bar semi-major axis
& $a_{\rm bar}\simeq1.3$--$2.3$ kpc under the suspected-bar interpretation; uncertain [7]
& $a_{\rm bar}\simeq4.8\,{\rm kpc}$ provisional deprojected estimate, conditional on the adopted disc geometry [TW]
& $a_{\rm bar}=3.18\pm0.32$ kpc [25]
& Unconstrained; the decomposition provides $r_{e,\rm bar}\simeq2.29$ kpc, not $a_{\rm bar}$ [47,TW] \\

Relative bar length
& Reported normalised bar length $r_{\rm bar,d}/r_e=0.28$ under the suspected-bar interpretation; definition-sensitive and not directly comparable to $a_{\rm bar}/h$ [8]
& Bar-to-disc-scale-length ratio $a_{\rm bar}/h=1.71^{+0.27}_{-0.16}$ [TW]
& $a_{\rm bar}/R_{\rm d}\simeq1.08$, derived from the tabulated bar semi-major axis and disc scale length [TW] 
& Unconstrained because $a_{\rm bar}$ is not reported \\

Corotation-to-bar ratio
& Unconstrained
& Unconstrained
&  $\mathcal{R}\equiv R_{\rm CR}/a_{\rm bar}=1.09\pm0.14$ [25]
&  Unconstrained \\

\hline
\multicolumn{5}{c}{%
\textbf{Stage III. Milky-Way-Compatible Nuclear State
($ {\rm MWA}\cap{\rm Isolated}\cap{\rm Barred}
\cap{\rm MW\mbox{-}compatible~nuclear~state} $)}}\\
\hline
& \textbf{NGC~3521} & \textbf{UGC~11487} & \textbf{NGC~2903} & \textbf{CGCG~047-026} \\
\hline

Supermassive black-hole mass
& $M_{\rm SMBH}=2.6\times10^{6}\,M_\odot$ from the $M_{\rm SMBH}$--S\'ersic-index relation [38]; $\log(M_{\rm SMBH}/M_\odot)=6.85\pm0.58$ from the spiral-arm pitch-angle relation [39]
&
$M_{\rm SMBH}\simeq6.3\times10^{6}\,M_\odot$ [$\log(M_{\rm SMBH}/M_\odot)=6.8\pm0.5$], inferred from the host stellar mass using the $M_{\rm SMBH}$--$M_\star$ relation; not a direct dynamical measurement  [20]
& $M_{\rm SMBH}\simeq3.0\times10^{6}\,M_\odot$ ($M_{\rm SMBH}$--$\sigma_\star$ estimate; not a direct dynamical measurement) [28]
& $M_{\rm BH}\sim(1.5$--$4.4)\times10^{7}\,M_\odot$; indirect estimates from the $M_{\rm BH}$--$M_{\rm bulge}$ and $M_{\rm BH}$--$\sigma_\star$ relations, respectively[TW] \\

Nuclear state
& Optical nuclear classification H\,{\sc ii}/LINER or LINER-like, depending on the adopted optical diagnostic [9,14]
& Star-forming in the BPT diagrams, with narrow emission lines and no Bowen-fluorescence or coronal-line evidence for persistent AGN activity [17,19]; the MIR-selected TDE-like flare WTP14adeqka is treated as a transient accretion event [19,20]
& H\,{\sc ii}/nuclear-starburst classification; no nuclear point-like X-ray source is detected, and there is no compelling evidence for a persistent AGN [18,31]
& LINER/LIER line ratios; $|{\rm EW}({\rm H}\alpha)|\simeq0.61$~\AA, consistent with a retired nucleus rather than a persistent AGN [TW] \\

X-ray Eddington ratio
& $L_{0.3-8\,{\rm keV}}/L_{\rm Edd}=2.2\times10^{-7}$ [38]
& $L_{\rm X}/L_{\rm Edd}\lesssim8.0\times10^{-4}$  [TW]
& $L_{0.5-8\,{\rm keV}}/L_{\rm Edd} \lesssim5.3\times10^{-7}$, based on the Chandra upper limit for an unresolved nuclear source and the adopted $M_{\rm SMBH}$--$\sigma_\star$ mass estimate [18,28,TW]
& Unconstrained; the weak optical emission may not be AGN-powered\\

\hline
\multicolumn{5}{c}{%
\textbf{Stage IV. Global Physical Constraints}}\\
\hline
& \textbf{NGC~3521} & \textbf{UGC~11487} & \textbf{NGC~2903} & \textbf{CGCG~047-026} \\
\hline

Total stellar luminosity
& $L_\star=(7.59\pm0.38)\times10^{10}\,L_\odot$ from panchromatic SED modelling [9]
& $L_\star=(5.36\pm0.27)\times10^{10}\,L_\odot$ from panchromatic SED modelling [19]
& $L_\star=(3.82\pm0.19)\times10^{10}\,L_\odot$ [TW]
& Unconstrained \\

Dust luminosity
& $L_{\rm dust}=(2.66\pm0.13)\times10^{10}\,L_\odot$ from panchromatic SED modelling [9]
& $L_{\rm dust}\simeq2.75\times10^{10}\,L_\odot$ [$\log(L_{\rm dust}/L_\odot)\simeq10.44$] [19]
& $L_{\rm dust}=(1.73\pm0.09)\times10^{10}\,L_\odot$ [TW]
& Unconstrained \\

Radio spectral index
& $\alpha_{\rm nt}=0.95\pm0.08$ over $1.4$--$10.5$~GHz, adopting $S_\nu^{\rm nt}\propto\nu^{-\alpha_{\rm nt}}$ [23]
& $\alpha\simeq0.5$ for the detected radio component; not treated as a homogeneous quiescent global-host spectral index [17]
& $\alpha_{\rm high}=0.91\pm0.01$ above the fitted break near 1.38~GHz, adopting $S_\nu\propto\nu^{-\alpha}$ [30]
& Unconstrained \\

Optical/stellar disc size
& $R_{25}=12.94$ kpc for the adopted distance $D=10.7$ Mpc [10]
& Isophotal optical radius, $R_{25}$, Unconstrained; $r_{e,\rm disc}=4.66^{+0.25}_{-0.47}$~kpc, derived from $r_{e,\rm disc}=1.678h$ [TW]
& $R_{25}=15.21$ kpc [10]
& Unconstrained \\

Stellar mass
& $M_\star=(5.99\pm0.30)\times10^{10}\,M_\odot$ from panchromatic SED modelling [9]
& $(2.01\pm0.15)\times10^{10}\,M_\odot$ [19]
& $M_\star=3.50\times10^{10}\,M_\odot$ [16], compared with $(1.36\pm0.21)\times10^{10}\,M_\odot$ from our homogeneous SED fit [TW]; the difference reflects SED-model, IMF, and decomposition systematics
& $M_\star\simeq4.8\times10^{10}\,M_\odot$; photometric estimate obtained by summing the disc, bar, and bulge stellar masses [TW] \\

Star-formation rate
& ${\rm SFR}=1.65\pm0.08\,M_\odot\,{\rm yr}^{-1}$ [9]
& ${\rm SFR}=2.92\pm0.15\,M_\odot\,{\rm yr}^{-1}$; ${\rm sSFR}=(1.45\pm0.13)\times10^{-10}~{\rm yr}^{-1}$ [19,TW]
& ${\rm SFR}=2.2\,M_\odot\,{\rm yr}^{-1}$ [15]; the homogeneous SED fit gives $1.47\pm0.09\,M_\odot\,{\rm yr}^{-1}$ [TW]
& Unconstrained \\

Halo mass
& $M_{200}=1.13\times10^{12}\,M_\odot$ for the constrained NFW fit, the Burkert fit gives $M_{200}\simeq3.0\times10^{11}\,M_\odot$ [11]
& Unconstrained
& $M_{\rm vir}\simeq(0.66$--$0.71)\times10^{12}\,M_\odot$ [29]
& Unconstrained \\

Characteristic rotation velocity
& $V_{\rm rot}\simeq227\,{\rm km\,s^{-1}}$ [10]
& $V_{\rm rot,W}\simeq149\,{\rm km\,s^{-1}}$ [TW]
& $V_{\rm rot}\simeq215\,{\rm km\,s^{-1}}$ [10]
& Unconstrained \\

Gas fraction
& $f_{\rm gas}=0.195$, including atomic and molecular gas and the factor 1.36 for helium [TW]
& Atomic-gas-based lower limit $f_{\rm gas}\gtrsim0.296$, including the factor 1.36 for helium; molecular gas is not included [TW]
& $f_{\rm gas}=0.228$, including atomic and molecular gas and the factor 1.36 for helium [TW]
& Unconstrained \\

Neutral hydrogen mass
& $M_{\rm HI}=8.02\times10^{9}\,M_\odot$ [13]
& $M_{\rm HI}=6.22\times10^{9}\,M_\odot$, derived from $F_{\rm HI}=3.9707~{\rm Jy\,km\,s^{-1}}$ [TW]
& $M_{\rm HI}=(4.8\pm0.3)\times10^{9}\,M_\odot$ [15]
& Unconstrained \\

Molecular gas mass
& $M_{\rm H_2}=2.65\times10^{9}\,M_\odot$ [13]
& Unconstrained
& $M_{\rm H_2}=(2.8\pm0.6)\times10^{9}\,M_\odot$ [26]
& Unconstrained \\

\hline
\multicolumn{5}{c}{%
\textbf{Stage V. Panchromatic SED Similarity}}\\
\hline
& \textbf{NGC~3521} & \textbf{UGC~11487} & \textbf{NGC~2903} & \textbf{CGCG~047-026} \\
\hline

Integrated SED
& Broad agreement with the Milky Way from the UV to radio/decametre wavelengths, with the principal residual differences occurring in the far-UV, near the far-IR dust peak, and in the centimetre-radio component [9,TW]
& Strong MW SED-shape similarity over $0.12$--$5\,\mu{\rm m}$ ($r=0.946$, $r_{\rm S}=0.905$, ${\rm median}|\Delta s|=0.272$, ${\rm RMS}(\Delta s)=0.624$), strongest at $1$--$5\,\mu{\rm m}$; MIR, FIR, and radio data are excluded because of upper limits and possible transient contamination [TW]
& Broad agreement with the Milky Way over the available UV--FIR range; the normalised SED slopes are strongly correlated, with $r=0.870$ and $r_{\rm S}=0.850$ [TW]
& No homogeneous UV-to-radio SED comparison with the Milky Way is currently available \\

\hline
\multicolumn{5}{c}{%
\textbf{Stage VI. Rotation-Curve Similarity}}\\
\hline
& \textbf{NGC~3521} & \textbf{UGC~11487} & \textbf{NGC~2903} & \textbf{CGCG~047-026} \\
\hline

Rotation-curve morphology
& The SPARC rotation curve reaches $V_{\rm rot,max}\simeq220~{\rm km\,s^{-1}}$ at $R\simeq5$~kpc, close to the characteristic Milky Way velocity scale. Its fitted outer slope, $\langle dV_{\rm rot}/dR\rangle=-0.68\pm0.15~{\rm km\,s^{-1}\,kpc^{-1}}$, is substantially shallower than the Milky Way estimates of $-1.81$ to $-1.90~{\rm km\,s^{-1}\,kpc^{-1}}$ [40--43,TW]
& No spatially resolved rotation curve is available. The unresolved double-horned GBT H\,{\sc i} profile gives $W_{mx}=257.94^{+1.96}_{-3.67}~{\rm km\,s^{-1}}$, $V_{\rm rot,W}=149.38^{+3.62}_{-3.34}~{\rm km\,s^{-1}}$, and $K=-0.049\pm0.012$; these quantities constrain the global profile but not $V(R)$, $V_{\rm flat}$, or the outer rotation-curve slope [TW]
& The resolved SPARC rotation curve reaches $V_{\rm rot,max}\simeq216~{\rm km\,s^{-1}}$ and subsequently declines. Over $6\leq R\leq20$~kpc, its fitted slope is $\langle dV_{\rm rot}/dR\rangle=-1.70\pm0.08~{\rm km\,s^{-1}\,kpc^{-1}}$, close to the Milky Way estimates of $-1.81$ to $-1.90~{\rm km\,s^{-1}\,kpc^{-1}}$ [40--43,TW]
& No spatially resolved rotation curve or reliable $V_{\rm flat}$ measurement is currently available \\

\hline
\multicolumn{5}{c}{%
\textbf{Stage VII. Evolutionary Fingerprints and Fossil Tracers}}\\
\hline
& \textbf{NGC~3521} & \textbf{UGC~11487} & \textbf{NGC~2903} & \textbf{CGCG~047-026} \\
\hline

Resolved chemical evolution
& Broken gas-phase oxygen-abundance gradient with $R_{\rm b,O/H}=4.9\pm0.5$~kpc; the effective oxygen yield lies within the empirical yield interval [33]
& No resolved radial abundance distribution is currently available
& $\nabla_{\rm O/H}=-0.338\pm0.035$~dex~$R_{25}^{-1}$, corresponding to $-0.0222\pm0.0023$~dex~kpc$^{-1}$ [10,TW]
& Unconstrained \\

Merger and assembly history
& The intermediate-age stellar population and low-surface-brightness outer structures are compatible with minor accretion or secular evolution, but no unique merger event has been established [24,44]
& Unconstrained; the absence of statistically significant global H\,{\sc i} flux asymmetry does not by itself determine the past merger or accretion history
& No evidence for a recent major merger; minor mergers, gas accretion, and secular evolution are favoured [16]
& Unconstrained \\

Chemo-dynamical signatures
& Old, intermediate-age, and young stellar components together with asymmetric central stellar line-of-sight velocity distributions [7,9,24]
& Unconstrained; no resolved stellar-population or stellar/gas chemo-dynamical analysis is currently available
& Old, metal-rich central population; the bar is dominated by $1$--$5$~Gyr stellar populations with metallicities similar to those of the outer disc [16]
& $D_n(4000)=2.003\pm0.016$ and $H\delta_A=-3.25\pm0.42$~\AA, indicating an old, predominantly quiescent central stellar population [TW] \\

Nuclear time-domain tracers
& No comparable time-domain nuclear event is included in the present assessment
& WTP14adeqka -- TDE-like flare with a multi-zone infrared dust response ($\tau=224\pm45$~d, $R_{\rm echo}=0.188\pm0.038$~pc, $L_{\rm IR,peak}\sim3\times10^{43}\,{\rm erg\,s^{-1}}$, $E_{\rm IR}\simeq(2.9$--$3.8)\times10^{51}$~erg); a single self-similar response is strongly disfavoured ($\Delta{\rm BIC}=281.7$) [19]
& No comparable time-domain nuclear event is included in the present assessment
& Unconstrained \\

Globular-cluster system
& No homogeneous constraints on the total population, specific frequency, metallicity distribution, or kinematics
& Unconstrained
& Insufficient homogeneous constraints on the total population, metallicity distribution, and kinematics
& Unconstrained \\

Circumgalactic medium and disc--halo cycling
& A slow-rotating anomalous H\,{\sc i} component contains $\sim1.5\times10^{9}\,M_\odot$, has a scale height of $\sim3.5$~kpc, and predominantly favours galactic-fountain cycling [12]
& Unconstrained
& Diffuse soft X-ray-emitting gas extends to approximately 5~kpc and is partly extraplanar; the extended CGM remains unconstrained [18]
& Unconstrained \\

Large-scale multiwavelength fossil tracers
& Diffuse low-surface-brightness stellar envelope with cloud-like and possible shell-like substructures compatible with past minor accretion [44]
& Delayed, spatially resolved non-relativistic radio emission consistent with an outflow--medium interaction; delayed jet escape following hydrodynamic alignment remains an alternative interpretation [17,48]
& Extended soft X-ray emission associated with starburst-driven disc--halo feedback, without an unambiguous large-scale wind morphology [18]
& Unconstrained \\

\hline
\multicolumn{5}{c}{\textbf{Integrated MWNeT Assessment}}\\
\hline
& \textbf{NGC~3521} & \textbf{UGC~11487} & \textbf{NGC~2903} & \textbf{CGCG~047-026} \\
\hline

Overall classification
& Strong MWNeT pathfinder candidate with broad agreement across environmental, global, panchromatic, and evolutionary diagnostics; the principal qualifications are the uncertain physical identification of the bar and the comparatively shallow decline of the outer rotation curve
& Promising time-domain-selected MWNeT candidate passing the isolation, structural, and nuclear-state stages; its global properties and UV--NIR SED are broadly Milky-Way-compatible, but its stellar mass is below the fiducial range and the panchromatic and resolved-kinematic assessments remain incomplete
& Strong, well-constrained MWNeT candidate with compatible environmental, structural, nuclear, global, panchromatic, resolved-kinematic, and evolutionary diagnostics; the principal qualifications are its small $B/T$, stellar-mass systematics, and incomplete GC/CGM constraints
& Promising early-stage candidate that passes the isolation, barred-disc, and nuclear-state stages but requires global SED, gas, chemical, and resolved-kinematic follow-up \\

\end{longtable}
\endgroup

\begingroup
\footnotesize
\noindent
\medskip

\textit{Reference notes.}

[1] \citet{deVaucouleurs1991RC3}, RC3 morphology;
[2] \citet{Buta2015Cat}, CVRHS/S$^4$G morphology;
[3] \citet{Das2003}, spiral and weak/bar-like structure of NGC~3521;
[4] \citet{Leroy2008}, exponential stellar-disc scale length of NGC~3521;
[5] \citet{Zhou2012}, $3.6\,\mu$m bulge--bar--disc decomposition of NGC~2903;
[6] \citet{Salo2015}, structural decomposition of NGC~3521;
[7] \citet{Zeilinger2001}, asymmetric stellar LOSVDs and bar-like interpretation of NGC~3521;
[8] \citet{Zurita2021}, homogeneous disc/bar measurements;
[9] \citet{Kompaniiets2026}, panchromatic modelling and global properties of NGC~3521;
[10] \citet{Pilyugin2023}, optical radii and radial gas-phase oxygen-abundance distributions;
[11] \citet{HessmanZiebart2011}, dark-halo modelling of NGC~3521;
[12] \citet{Elson2014}, H\,{\sc i} rotation curve of NGC~3521;
[13] \citet{Grisdale2017}, atomic and molecular gas masses of NGC~3521;
[14] \citet{Kompaniiets2025}, environment and nuclear classification of NGC~3521;
[15] \citet{Irwin2009}, environment, SFR, H\,{\sc i}, and rotation curve of NGC~2903;
[16] \citet{Carrillo2020}, structure, stellar populations, and evolutionary history of NGC~2903;
[17] \citet{Golay2026}, nuclear classification, radio continuum, and resolved outflow of UGC~11487;
[18] \citet{Yukita2012}, Chandra nuclear-starburst and extended-hot-gas analysis of NGC~2903;
[19] \citet{Izviekova2026}, MIR variability and multiwavelength SED modelling of UGC~11487;
[20] \citet{Masterson2024}, MIR-selected TDE candidates and SMBH-mass estimate for UGC~11487;
[21] \citet{Karachentseva1973}, KIG catalogue;
[22] \citet{Karachentseva2010}, 2MIG catalogue and isolation criterion;
[23] \citet{Tabatabaei2017}, non-thermal radio spectral index of NGC~3521;
[24] \citet{Coccato2018}, bulge--disc decomposition and inner structure of NGC~3521;
[25] \citet{Salak2019}, molecular-gas kinematics and bar pattern speed of NGC~2903;
[26] \citet{Muraoka2016}, molecular-gas mass of NGC~2903;
[27] \citet{Karachentsev2015LeoSpur}, Local-Volume environment of NGC~2903;
[28] \citet{She2017ChandraCatalog}, $M_{\rm BH}$--$\sigma_\star$ estimate for NGC~2903;
[29] \citet{SylosLabini2024}, rotation-curve halo modelling of NGC~2903;
[30] \citet{Chyzy2018}, global radio-continuum spectrum of NGC~2903;
[31] \citet{Karachentseva2011Companions}, search for faint companions around 2MIG galaxies using $R_p<500$ kpc and $|\Delta V|<500~{\rm km\,s^{-1}}$;
[32] \citet{Regan2006}, stellar-disc scale length of NGC~3521 derived from the Spitzer $3.6\,\mu$m exponential-disc fit;
[33] \citet{Pilyugin2025}, abundance gradient of NGC~3521;
[34] \citet{Inman2026}, multi-component radiative-transfer modelling of NGC~3521, including inner, main, and outer stellar-disc components;
[35] \citet{Kourkchi2017}, cosmic-web location of NGC~3521;
[36] \citet{Karachentsev2022NGC3115}, dwarf-galaxy candidates in the satellite environment of NGC~3521 identified with DESI imaging;
[37] \citet{Muller2025}, dwarf-galaxy candidates in the satellite environment of NGC~3521 identified with Subaru/HSC; the degree of overlap between [36] and [37], and the physical association of the candidates with NGC~3521, require confirmation through distance or velocity measurements;
[38] \citet{Zhang2009}, SMBH mass of NGC~3521 from the S\'ersic-index relation;
[39] \citet{Davis2014}, SMBH mass of NGC~3521 from the spiral-arm pitch angle;
[40] \citet{Fedorov2026}, Milky Way two-dimensional rotation curve;
[41] \citet{Ou2024}, Milky Way rotation curve;
[42] \citet{deBlok2008}, resolved rotation curve of NGC~3521;
[43] \citet{Lelli2016}, SPARC rotation-curve compilation;
[[44] \citet{MartinezDelgado2010}, deep imaging and diffuse stellar substructure around NGC~3521;
[45] \citet{Paturel2003} HYPERLEDA Catalog of galaxies;
[46] \citet{Carlsten2022} satellite system of NGC 3521;
[47] \citep{Kruk2018} CGCG~047-026 multiband \(ugriz\) decomposition;
[48] \citet{TeboulMetzger2023}, delayed jet escape following prolonged hydrodynamic alignment in tidal disruption events;
[TW] This work.
\par
\endgroup

\end{document}